\documentclass[twocolumn,prb,10pt,aps,floatfix]{revtex4-1}
\usepackage[english]{babel}
\usepackage{graphicx}
\usepackage[pdftex]{epsfig}
\usepackage{ragged2e}
\usepackage[caption=false]{subfig}
\usepackage{amsmath,amssymb}
\usepackage{times}
\usepackage[colorlinks,linkcolor=blue,citecolor=blue,urlcolor=blue]{hyperref}
\usepackage{color}
\usepackage[table,xcdraw]{xcolor}
\usepackage{epstopdf}
\usepackage{physics}
\usepackage[export]{adjustbox}
\usepackage{feynmf}
\usepackage{dsfont}

\newcommand{\lau}[1]{\textcolor{blue}{Laura: #1}}

\newcommand{\be}{\begin{equation}}
\newcommand{\ee}{\end{equation}}

\newcommand{\bee}{\begin{equation*}}
\newcommand{\eee}{\end{equation*}}

\begin{document}
\title{Numerical treatment of spin systems with unrestricted spin length $S$: A functional renormalization group study}
\author{M. L. Baez$^{1, 2}$}
\author{J. Reuther$^{1, 2}$}
\affiliation{$^1$Dahlem Center for Complex Quantum Systems, Freie Universit\"{a}t Berlin, Germany}
\affiliation{$^2$Helmholtz-Zentrum f\"{u}r Materialien und Energie, Berlin, Germany} 

\date{\today}

\begin{abstract}
We develop a generalized pseudo-fermion functional renormalization group (PFFRG) approach that can be applied to arbitrary Heisenberg models with spins ranging from the quantum case $S=1/2$ to the classical limit $S\rightarrow\infty$. Within this framework, spins of magnitude $S$ are realized by implementing $M=2S$ copies of spin-1/2 degrees of freedom on each lattice site. We confirm that even without explicitly projecting onto the highest spin sector of the Hilbert space, ground states tend to select the largest possible local spin magnitude. This justifies the average treatment of the pseudo fermion constraint in previous spin-1/2 PFFRG studies. We apply this method to the antiferromagnetic $J_1$-$J_2$ honeycomb Heisenberg model with nearest neighbor $J_1>0$ and second neighbor $J_2>0$ interactions. Mapping out the phase diagram in the $J_2/J_1$-$S$ plane we find that upon increasing $S$ quantum fluctuations are rapidly decreasing. In particular, already at $S=1$ we find no indication for a magnetically disordered phase. In the limit $S\rightarrow\infty$, the known phase diagram of the classical system is exactly reproduced. More generally, we prove that for $S\rightarrow\infty$ the PFFRG approach is identical to the Luttinger-Tisza method.  
\end{abstract}
\maketitle

%====================================================================
\section{Introduction}

Frustrated quantum spin systems harbor a plethora of fascinating ground-state phenomena arising in a situation when quantum fluctuations are strong enough to melt magnetic long-range order. A prominent class of exotic quantum phases are spin liquids\cite{anderson73, Balents10} of various different types, associated with novel concepts\cite{savary16} such as long-range entanglement, topologically protected degeneracies or fractional quasi-particle excitations. While the traditional recipe for maximizing the effects of quantum fluctuations primarily involves spins of the smallest magnitude $S=1/2$, interesting spin phases can likewise occur in the extreme opposite limit of classical spins with $S\rightarrow\infty$. For example, this limit is approximately realized in classical spin-ice materials\cite{bramwell01} which are characterized by an extensive ground state degeneracy\cite{ramirez99} and effective monopole excitations.\cite{castelnovo08} Furthermore, there is a growing number of spin systems where novel types of quantum phases appear at a specific intermediate value of $S$ (see e.g. Refs~\onlinecite{affleck87,yao09,wei11,picot15}).

Due to the strongly correlated nature of quantum spin systems, detecting the aforementioned phenomena within numerical approaches is generally a very difficult task. While there exists a number of powerful approaches to treat the spin-1/2 case, each method is also characterized by certain weaknesses. For example, exact diagonalization is free of any errors for small spin clusters, but extrapolating the physical properties to the thermodynamic limit can be challenging. Similarly, quantum Monte Carlo\cite{sandvik91,ceperley77,reger88} is -- up to statistical errors -- exact in non-frustrated coupling scenarios, however, the frustrated case is generally not accessible due to the sign problem. The density matrix renormalization group (DMRG) method\cite{white92,schollwoeck05} has proven to be very powerful in 1D and sometimes also in 2D\cite{depenbrock12,bauer14} but 3D spin systems seem to be out of reach for this approach. The classical case $S\rightarrow\infty$ can likewise be challenging and there is a separate class of approaches such as the Luttinger-Tisza (LT) method\cite{luttinger46,luttinger51} or classical Monte Carlo techniques which have proven powerful in this situation. However, since interesting ground-state phases can occur at all spin lengths, numerical methods that can be easily tuned between the extreme quantum and classical limits are highly desirable.

In this article, we propose a numerical scheme based on the PFFRG approach that can be applied to arbitrary spin lengths $S$ within the same methodological framework. For $S=1/2$, this technique has already been used to investigate frustrated spin systems, yielding an accurate description of the interplay between magnetically ordered and disordered phases.\cite{reuther10,balz16,iqbal15,iqbal16,buessen16,hering16,reuther11,reuther11_2} In particular, the strength of this approach lies in its flexibility, allowing for complex coupling scenarios such as longer-range frustrated interactions on complicated lattices\cite{balz16,iqbal15} (including 3D systems\cite{iqbal16,buessen16}) as well as anisotropic couplings.\cite{hering16,reuther11,iqbal162} Concerning its limitations, the current implementation of the PFFRG can not resolve all possible magnetically disordered phases as it systematically misses certain types of (three body) spin correlations which are, e.g., important for the description of chiral spin-liquid phases. Here, we further extend the flexibility of the PFFRG method by generalizing it to arbitrary spin magnitudes. Proposing a scheme where multiple copies of spin-1/2 degrees of freedom are considered on each lattice site, we are able to investigate spin systems between the $S=1/2$ quantum case and the classical limit $S\rightarrow\infty$ including all possible intermediate values. Most importantly, the required modifications for varying the spin length turn out to be surprisingly simple.

\begin{figure}
%\centering
\includegraphics[scale=0.9,angle=0]{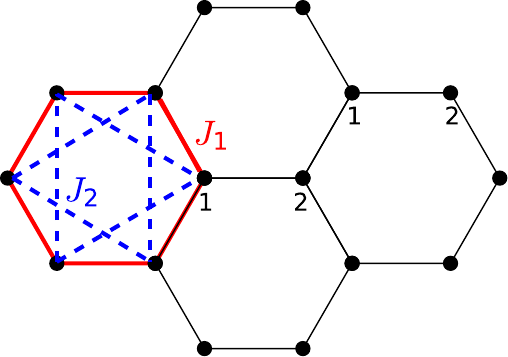}
\caption{Illustration of the honeycomb lattice where $J_1$ nearest neighbor ($J_2$ second neighbor) interactions are highlighted by red (dashed blue) lines. The two sublattices are indicated by numbers and the nearest neighbor distance is assumed to be one throughout the paper.}
\label{lattice}
\end{figure}

As a first test we apply this scheme to the antiferromagnetic $J_1$-$J_2$ Heisenberg model on the honeycomb lattice with first (second) neighbor interactions $J_1$ ($J_2$), see Fig.~\ref{lattice}. Due to the frustrating effect of the $J_2$ coupling, the system shows rich magnetic behavior as a function of $S$ and $J_2/J_1$, where the spin-1/2 case has attracted particular attention. While the system remains antiferromagnetically ordered up to $J_2/J_1\approx0.2$, an abundance of numerical studies for $S=1/2$ indicate an intermediate magnetically disordered phase above this value.\cite{reuther11_2,mulder10,oitmaa11,albuquerque11,fouet01,mosadeq11,yu14,zhu13,ganesh13,gong13,zhang13,bishop13,li12,mezzacapo12,bishop12} The precise nature of this phase is still debated, but there is growing numerical evidence that it might again be split up into a potential plaquette valence bond solid at smaller $J_2/J_1$ and a staggered dimer crystal phase at larger $J_2/J_1$.\cite{mosadeq11,zhu13,ganesh13,bishop13} Concerning the opposite limit $S\rightarrow\infty$\cite{rastelli79} it has early been realized that above the classical antiferromagnetic phase (which is stable up to $J_2/J_1=1/6$) the system features a continuous set of degenerate incommensurate spiral ground states\cite{mulder10,katsura86} where quantum fluctuations at infinitesimal $1/S$ select a finite subset of these states.\cite{mulder10} Even though the phase diagrams at small and large $S$ differ considerably, raising questions about the magnetic properties at intermediate spin lengths, systematic studies with unrestricted $S$ are rather rare so far. Numerical investigations based on coupled cluster and DMRG approaches indicate that a small non-magnetic phase might survive in the $S=1$ case.\cite{gong15,li16} There is also growing experimental interest in these systems, stemming from honeycomb materials with different spins, such as $\textrm{Bi}_3\textrm{Mn}_4\textrm{O}_{12}(\textrm{NO}_3)$ hosting spin-3/2 Mn$^{4+}$ ions\cite{onishi12,smirnova09,okubo10} or BaNi$_2$V$_2$O$_8$ based on spin-1 Ni$^+$ ions.\cite{rogado02,krug03}

The spin-1/2 antiferromagnetic $J_1$-$J_2$ Heisenberg model on the honeycomb lattice has previously been studied with PFFRG\cite{reuther11_2}, showing a magnetically disordered phase at $J_2/J_1\approx 0.15\ldots0.6$. Probing this regime with respect to the formation of different types of valence-bond crystals, strong staggered dimer responses are found near the upper boundary of this phase, in agreement with other numerical studies.\cite{mosadeq11,zhu13,ganesh13,gong13,zhang13,bishop13} Near the lower phase boundary ($J_2\sim0.2$) the PFFRG dimer responses are small, possibly pointing at the existence of a spin liquid phase. At larger spin lengths studied in this work, we find that the phase diagram quickly resembles the classical one. In particular, already at $S=1$, PFFRG shows no indication of a magnetically disordered phase. Instead, the system exhibits two regimes with different types of incommensurate magnetic spiral phases. For $S=3$ the magnetic phase diagram is almost indistinguishable from the one at $S\rightarrow\infty$ except that quantum fluctuations select specific states out of the continuous set of degenerate classical states. This selection is found to be in agreement with earlier semiclassical studies.\cite{mulder10} For $S\rightarrow\infty$ the PFFRG equations can be solved analytically and we exactly reproduce the known classically ordered states. More generally, we demonstrate that for arbitrary lattices the PFFRG becomes identical to the LT method in this limit. 

The paper is structured as follows: Sec.~\ref{sec:method} introduces the PFFRG method, wherein Sec.~\ref{sec:meth_half} first reviews the standard spin-1/2 case. Afterwards, Sec.~\ref{sec:ls} describes the extension of this technique to arbitrary spin quantum numbers. In Sec.~\ref{sc:local} we test its accuracy by considering the effects of additional level repulsion terms. We apply the approach to the antiferromagnetic $J_1$-$J_2$ Heisenberg model on the honeycomb lattice in Sec.~\ref{sec:j1j2}, obtaining a phase diagram in the $J_2/J_1$-$S$ plane, see Sec.~\ref{sec:phd}. We particularly focus on the classical case $S\rightarrow\infty$ (Sec.~\ref{sec:largeS}) and show that an analytical solution is possible in this limit. The paper ends with a conclusion and discussion in Sec.~\ref{sec:conclusion}.

%====================================================================
\section{Method}
\label{sec:method}
\subsection{Introduction to the general PFFRG procedure}
\label{sec:meth_half}
The PFFRG method for quantum spin systems is a variant of the more general FRG framework\cite{wetterich93,metzner12,platt13} which is used, e.g., to investigate Hubbard-like fermionic systems.\cite{honerkamp01,zanchi00} So far the PFFRG has been applied to spin-1/2 Heisenberg models in two and three dimensions\cite{reuther10,reuther11_2,balz16,iqbal15,iqbal16,buessen16} as well as to spin-anisotropic models in two dimensions.\cite{hering16,reuther11,iqbal162} Taking into account interaction processes in various different coupling channels, this approach yields a rather accurate distinction between magnetically ordered and disordered phases even in strongly frustrated scenarios. Before we explain our implementation of a spin-$S$ generalization of the PFFRG, we first briefly review the spin-1/2 case. Particularly, we consider Heisenberg models of the form
\be
\mathcal{H} =  \sum_{(ij)}J_{ij} \mathbf{S}_i\cdot\mathbf{S}_j\,,
\label{hamiltonian}
\ee
where $i$, $j$ are the sites of an arbitrary lattice (later we will consider the honeycomb lattice). Furthermore, the symbol $(i,j)$ denotes pairs of sites (each pair is summed over only once) and $J_{ij}$ are the exchange couplings.

The PFFRG method starts with expressing the spin-1/2 operators in terms of pseudo fermions\cite{abrikosov65}
\be
S_{i}^{\mu}=\frac{1}{2}\sum_{\alpha \beta}f_{i\alpha}^{\dagger}\sigma_{\alpha\beta}^{\mu}f_{i\beta}\,,\label{representation}
\ee
where $\alpha,\beta = \uparrow,\downarrow$ denote spin indices, $f^{(\dagger)}_{i\alpha}$ are fermionic annihilation (creation) operators on site $i$ and $\sigma^\mu$ ($\mu=x,y,z$) represent the Pauli matrices. While this representation fulfills the correct angular momentum algebra of spin operators, the introduction of pseudo fermions is associated with an enlargement of the Hilbert space. Denoting the vector space of an arbitrary angular momentum operator $\mathbf{L}$ by $\mathds{V}^L$, the pseudo fermionic representation extends the spin-1/2 vector space $\mathds{V}^{1/2}$ according to
\be
\mathds{V}^{1/2}\rightarrow\mathds{V}^{0}\oplus\mathds{V}^0\oplus\mathds{V}^{1/2}\;,
\ee
where the symbol $\oplus$ denotes a direct sum. One finds that the physical spin-1/2 subspace $\mathds{V}^{1/2}$ is represented by the two basis states $|f_{i\uparrow}^\dagger f_{i\uparrow},f_{i\downarrow}^\dagger f_{i\downarrow}\rangle=|1,0\rangle$ and $|0,1\rangle$ while the two spin-0 subspaces $\mathds{V}^0$ are given by the states $|0,0\rangle$ and $|1,1\rangle$. In order to treat the original spin-1/2 model one needs to project out possible spurious admixtures from the unphysical spin-0 states. While this is in general a non-trivial problem, the situation simplifies considerably at zero temperature. This can be seen by noting that non or doubly occupied spin-0 sites are equivalent to vacancies in the spin lattice. To create such a vacancy (e.g. via a fermion number fluctuation on a particular site) the binding energy of a spin to its environment needs to be overcome. It therefore appears plausible that the ground state of the fermionic system lies entirely in the physical spin-1/2 sector and that unphysical occupations are gapped excitations with an energy on the order of the exchange couplings. In Sec.~\ref{sc:local} we will show that this is indeed the case, proving that at $T=0$ the pseudo fermion constraint is automatically fulfilled without any further methodological adjustments. 

The introduction of the pseudo fermions enables us to use diagrammatic many-body techniques such as FRG. Without any quadratic terms in the pseudo-particle Hamiltonian the bare fermionic propagator in Matsubara space is simply given by
\be
G_0(1';1)=\frac{1}{i\omega_1}\delta(\omega_1-\omega_{1'})\delta_{i_{1'} i_1}\delta_{\alpha_1 \alpha_{1'}}\;,\label{g0}
\ee
where the index ``$1=\{\omega_1,i_1,\alpha_1\}$'' denotes a multi index containing the frequency variable $\omega_1$, the site index $i_1$ and the spin index $\alpha_1$. Also note that in the zero temperature limit considered here, the discrete Matsubara frequencies become continuous. The diagonal structure of Eq.~(\ref{g0}) in the frequency, site and spin variables is due to energy conservation, absence of any fermion hopping in the Hamiltonian, and isotropy in spin space, respectively.

Within PFFRG, the singularity of the propagator at $\omega=0$ is regularized by introducing an artificial infrared cutoff $\Lambda$ implemented via a Heavyside step-function,
\be
G_0^{\Lambda}(1';1) =\Theta(|\omega_1|-\Lambda)G_0(1';1)\;.
\ee 
This modification generates a $\Lambda$ dependence of all fermionic one-particle irreducible $m$-particle vertex functions $\Gamma_m^\Lambda$ such as the self energy $\Sigma^\Lambda(1';1)\equiv\Gamma_1^\Lambda(1';1)$ and the two-particle vertex $\Gamma^\Lambda(1',2';1,2)\equiv\Gamma_2^\Lambda(1',2';1,2)$. Following the standard FRG framework\cite{metzner12,platt13}, this dependence can be described by an infinite hierarchy of coupled integro-differential equations where the scale derivative $d\Gamma_m^\Lambda/d\Lambda$ couples to all vertices $\Gamma^\Lambda_n$ with $n=1,2,\ldots,m,m+1$. The equations for the self energy and the two-particle vertex take the form
\begin{align}\label{eq:FirstFlow}
 \frac{d}{d\Lambda}\Sigma^{\Lambda}\left(1';1\right)&=-\frac{1}{2\pi}\sum\limits_{2'\,2}\Gamma^{\Lambda}\left( 1',2';1,2\right)S^{\Lambda}\left(2,2'\right), \\ \frac{d}{d\Lambda}\Gamma^{\Lambda}\left(1',2';1,2\right)&=\frac{1}{2\pi}  \sum\limits_{3'\,3} \Gamma_3^\Lambda \left( 1',2',3';1,2,3\right) S^{\Lambda}\left(3,3'\right) \nonumber \\
 +\frac{1}{2\pi}\sum\limits_{3'\,3\,4'\,4}\:&\Big[ \Gamma^{\Lambda}\left( 1',2';3,4\right) \Gamma^{\Lambda}\left( 3',4';1,2\right) \nonumber \\ -\Gamma^{\Lambda}( 1',4';1&,3) \Gamma^{\Lambda}\left( 3',2';4,2\right) - \left (3'\leftrightarrow 4', 3 \leftrightarrow 4 \right ) \nonumber \\ +\Gamma^{\Lambda}( 2',4';1&,3) \Gamma^{\Lambda}\left( 3',1';4,2\right) + \left (3'\leftrightarrow 4', 3 \leftrightarrow 4 \right ) \Big ] \nonumber \\ \label{eq:SecondFlow} \times G^{\Lambda}(3,3')&S^{\Lambda}(4,4')\;,
\end{align}
where sums stand for $\Sigma_1\equiv\int_{\omega_1}d\omega_1\sum_{i_1}\sum_{\alpha_1=\uparrow,\downarrow}$ and $\Gamma_3^\Lambda$ is the three particle vertex. Furthermore, $G^\Lambda=[(G_0^\Lambda)^{-1}-\Sigma^\Lambda]^{-1}$ denotes the fully dressed propagator and 
\be
\label{eq:SingleScale}
 S^{\Lambda}=G^{\Lambda} \frac{d}{d\Lambda}\left( G^{\Lambda}_0 \right )^{-1} G^{\Lambda}
\ee
is the so-called single-scale propagator.

For a numerical evaluation of these equations, the infinite hierarchy needs to be truncated. The most straightforward truncation scheme amounts to treating the three-particle vertex $\Gamma^\Lambda_3$ as zero. This, however, leads to an insufficient feedback of the self energy into the two-particle vertex flow such that all results effectively remain on a classical level. Particularly, quantum fluctuations needed for the description of magnetically disordered phases are almost completely neglected within such a scheme.\cite{reuther10} The key improvement is achieved by the so-called Katanin truncation\cite{katanin04} which neglects $\Gamma^\Lambda_3$ in Eq.~(\ref{eq:SecondFlow}) but at the same time replaces the single-scale propagator by
\be
 \label{eq:Katanin}
 S^{\Lambda}\longrightarrow -\frac{d}{d\Lambda}G^{\Lambda}=S^{\Lambda}-\left(G^{\Lambda}\right)^2\frac{d}{d\Lambda}\Sigma^{\Lambda}\;.
\ee
This scheme effectively takes into account a certain subset of three-particle vertex contributions in Eq.~(\ref{eq:SecondFlow}). Most importantly, the modified single-scale propagator is given by the total derivative $-\frac{d}{d\Lambda}G^{\Lambda}$, such that the complete feedback of the self energy into the two-particle vertex is always ensured within the Katanin truncation. Since the self energy accounts for a finite pseudo fermion damping, this feedback is essential for the proper description of quantum fluctuations generating magnetically disordered phases.

The Katanin scheme reduces the FRG equations to a closed set which can be solved numerically. The initial conditions are usually taken in the limit $\Lambda\rightarrow\infty$ where the free propagator vanishes identically. Hence, the only finite vertex function at $\Lambda\rightarrow\infty$ is the bare two-particle vertex given by
\begin{align}
\Gamma^\infty(1',2';1,2)=&\frac{1}{4}J_{i_1 i_2}\sigma^\mu_{\alpha_{1'}\alpha_1}\sigma^\mu_{\alpha_{2'}\alpha_2}\delta_{i_{1'} i_1}\delta_{i_{2'} i_2}\notag\\
&\times\delta(\omega_1+\omega_2-\omega_{1'}-\omega_{2'})\notag\\
&-(\omega_1\leftrightarrow\omega_2,i_1\leftrightarrow i_2,\alpha_1\leftrightarrow\alpha_2)\;,\label{initial}
\end{align}
where the factor $\sim1/4\sigma^\mu\sigma^\mu$ results from the pseudo fermion representation (\ref{representation}) and a sum over $\mu$ is implicitly assumed. The last line guarantees that the fermionic antisymmetry condition under the exchange of variables $1\leftrightarrow2$ or $1'\leftrightarrow2'$ is fulfilled. Further note that due to the absence of any quadratic fermionic terms in the Hamiltonian, the self energy always vanishes identically at $\Lambda\rightarrow\infty$.

The flow equations can be brought into a more convenient form by exploiting the special site index structure of the two-particle vertex. Since all propagators $G^\Lambda(1',1)$, $S^\Lambda(1',1)$ are diagonal in $i_{1'}$, $i_1$ the spatial dependence of $\Gamma^\infty(1',2';1,2)$ as indicated in Eq.~(\ref{initial}) is retained to all levels of diagrammatic approximations. This means that for each diagrammatic contribution with site indices $i_{1'}$ and $i_{2'}$ on two external fermion lines, the other two indices must either be given by $i_1=i_{1'}$, $i_2=i_{2'}$ or $i_1=i_{2'}$, $i_2=i_{1'}$. The spatial dependence of $\Gamma^\Lambda(1',2';1,2)$ can therefore be parametrized as
\begin{align}
\Gamma^\Lambda(1',2';1,2)=&\tilde\Gamma_{i_1i_2}^\Lambda(1',2';1,2)\delta_{i_{1'} i_1}\delta_{i_{2'} i_2}\notag\\
&\times\delta(\omega_1+\omega_2-\omega_{1'}-\omega_{2'})\notag\\
&-(\omega_1\leftrightarrow\omega_2,i_1\leftrightarrow i_2,\alpha_1\leftrightarrow\alpha_2)\;,\label{parametrize}
\end{align}
where the new vertex $\tilde\Gamma^\Lambda$ fulfills the condition $\tilde\Gamma_{i_1i_2}^\Lambda(1',2';1,2)=\tilde\Gamma_{i_2i_1}^\Lambda(2',1';2,1)$. Note that the multi index ``1'' appearing in the argument of $\tilde\Gamma^\Lambda$ only stands for the frequency $\omega_1$ and the spin $\alpha_1$ while the site indices are written as a subscript. Furthermore, the $\delta$-function in the frequencies in Eq.~(\ref{parametrize}) guarantees that energy is conserved. The diagonal structure of the self energy in the frequency, site and spin variables allows us to write
\be
\Sigma^\Lambda(1';1)\equiv \Sigma^\Lambda_{i_1}(\omega_1)\delta(\omega_1-\omega_{1'})\delta_{i_{1'} i_1}\delta_{\alpha_1 \alpha_{1'}}\;,\label{parametrize_sigma}
\ee
and equivalently for $G^\Lambda(1',1)$ and $S^\Lambda(1',1)$. Inserting Eqs.~(\ref{parametrize}), (\ref{parametrize_sigma}) into Eqs.~(\ref{eq:FirstFlow}), (\ref{eq:SecondFlow}) and omitting the three-particle vertex yields
\begin{widetext}
\be
\frac{d}{d\Lambda}\Sigma_{i_1}^{\Lambda}\left(\omega_1\right)=\frac{1}{2\pi}\sum\limits_{2}\Big[-\sum_j\tilde\Gamma_{i_1j}^{\Lambda}(1,2;1,2)S_j^{\Lambda}(\omega_2)+\tilde\Gamma_{i_1i_1}^{\Lambda}(1,2;2,1)S_{i_1}^{\Lambda}(\omega_2)\Big]\;,\label{FRG_sigma}
\ee
\begin{align}
\frac{d}{d\Lambda}\tilde\Gamma_{i_1i_2}^{\Lambda}(1',2';1,2)
=&\frac{1}{2\pi}\sum_{3\,4}\Big[\tilde\Gamma_{i_1i_2}^{\Lambda}(1',2';3,4)\tilde\Gamma_{i_1i_2}^{\Lambda}(3,4;1,2)P_{i_1i_2}^\Lambda(\omega_3,\omega_4)\notag\\
&-\sum_j\tilde\Gamma_{i_1j}^{\Lambda}(1',4;1,3)\tilde\Gamma_{ji_2}^{\Lambda}(3,2';4,2)P_{jj}^\Lambda(\omega_3,\omega_4)
+\tilde\Gamma_{i_1i_2}^{\Lambda}(1',4;1,3)\tilde\Gamma_{i_2i_2}^{\Lambda}(3,2';2,4)P_{i_2i_2}^\Lambda(\omega_3,\omega_4)\notag\\
&+\tilde\Gamma_{i_1i_1}^{\Lambda}(1',4;3,1)\tilde\Gamma_{i_1i_2}^{\Lambda}(3,2';4,2)P_{i_1i_1}^\Lambda(\omega_3,\omega_4)
+\tilde\Gamma_{i_1i_2}^{\Lambda}(4,2';1,3)\tilde\Gamma_{i_1i_2}^{\Lambda}(1',3;4,2)P_{i_2i_1}^\Lambda(\omega_3,\omega_4)\Big]\,.\label{FRG_gamma}
\end{align}
\end{widetext}
Here, we have defined $P^\Lambda$ as a term containing all internal fermion lines, i.e.,
\be
P^\Lambda_{i_1 i_2}(\omega_1,\omega_2)=G_{i_1}^\Lambda(\omega_1)S_{i_2}^\Lambda(\omega_2)+G_{i_2}^\Lambda(\omega_2)S_{i_1}^\Lambda(\omega_1)\,.
\ee
The initial conditions for $\tilde\Gamma^\Lambda$ take the form
\be
\tilde\Gamma^\infty_{i_1 i_2}(1',2';1,2)=\frac{1}{4}J_{i_1 i_2}\sigma^\mu_{\alpha_{1'}\alpha_1}\sigma^\mu_{\alpha_{2'}\alpha_2}\;.\label{initial2}
\ee

\begin{figure*}
\includegraphics[scale=0.8,angle=0]{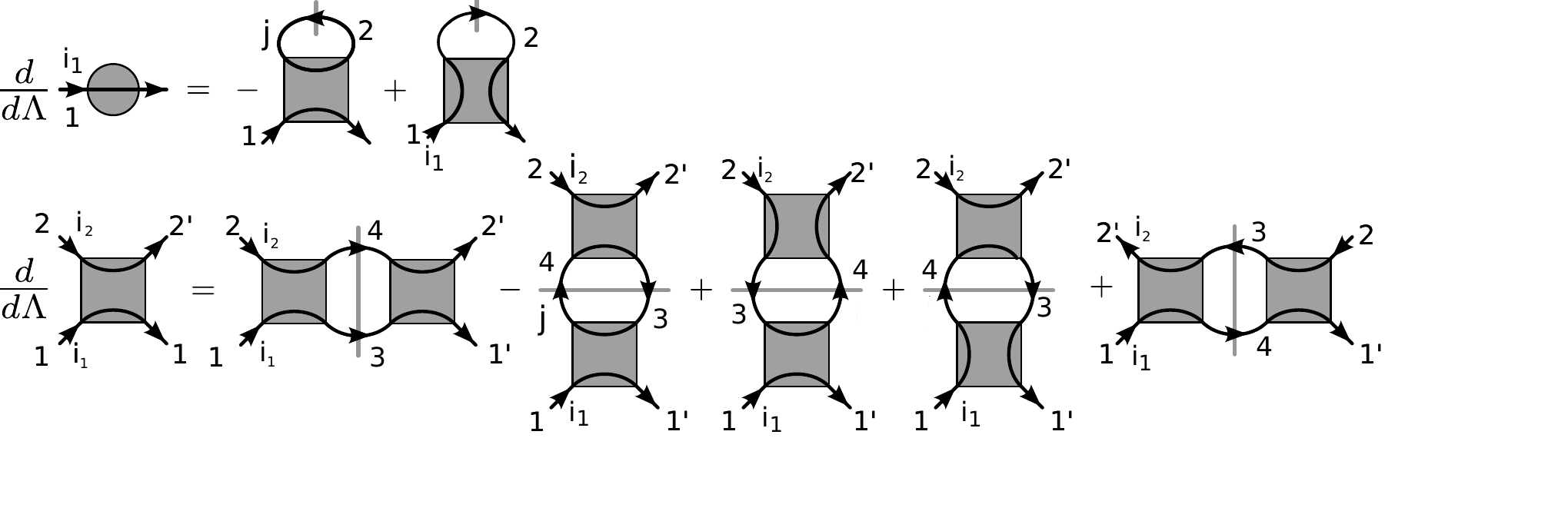}
\caption{Diagrammatic illustration of the PFFRG flow equations for the self energy and the two-particle vertex, see also Eqs.~(\ref{FRG_sigma}) and (\ref{FRG_gamma}). The gray lines crossing two fermion propagators denote the term $P^\Lambda_{i_1 i_2}(\omega_1,\omega_2)=G_{i_1}^\Lambda(\omega_1)S_{i_2}^\Lambda(\omega_2)+G_{i_2}^\Lambda(\omega_2)S_{i_1}^\Lambda(\omega_1)$ while slashes crossing only one line are the single scale propagators. Site indices $i_1$, $i_2$, $j$ illustrate the real-space structure of the flow equations. The five terms on the right-hand side of the second equation are the particle-particle channel, the RPA term, two vertex correction terms and the crossed particle hole term in the same order as they appear in Eq.~(\ref{FRG_gamma}).}
\label{FRGeq}
 \end{figure*}

The five terms on the right-hand side of Eq.~(\ref{FRG_gamma}) can be easily distinguished according to their site-index structure, as illustrated in Fig.~\ref{FRGeq}. The first term is a particle-particle term that generates ladder-type diagrams where the fermion lines have the same orientation (see arrows in Fig.~\ref{FRGeq}). The second term is special as it contains an internal closed fermion loop associated with a site summation. This term sums up RPA diagrams and will play an important role in the spin-$S$ generalization described below. Most importantly, this is the only term in the PFFRG equations where the vertex evolution $\frac{d}{d\Lambda}\tilde\Gamma^\Lambda_{i_1i_2}$ does not only couple to the local vertex $\tilde\Gamma_{ii}^\Lambda$ or to itself, but also to any other vertex $\tilde\Gamma_{i_1j}^\Lambda$ and $\tilde\Gamma_{ji_2}^\Lambda$. As a consequence, the RPA term generates long-range correlations between spins. The third and fourth terms in Eq.~(\ref{FRG_gamma}) are referred to as vertex corrections and the fifths term is the crossed particle-hole channel summing up ladder diagrams with fermion lines of opposite orientation. In general, the non-local nature of the RPA term is responsible for the formation of magnetic long-range order. On the other hand, the ladder diagrams induce a strong short-range binding between nearby spins leading to spin-singlet formation and to an effective non-magnetic resonating-valence bond description.

To numerically solve the PFFRG equations, several further steps of manipulating Eq.~(\ref{FRG_gamma}) need to be performed, such as parametrizing the spin dependences. In particular, a closed set of equations is only obtained when approximating the continuous frequency variables by a discrete grid. For our results below we use a combination of a linear and logarithmic mesh consisting of 40 discrete values for each frequency variable. Furthermore, the spatial dependence of the vertex functions $\tilde\Gamma^\Lambda_{i_1i_2}$ is approximated by only taking into account vertices where the distance between sites $i_1$ and $i_2$ does not exceed a maximal length which we choose to be 10 nearest neighbor lattice spacing. For a more detailed description of the PFFRG implementation we refer the reader to Ref.~\onlinecite{reuther10}. 

The two-particle vertex is directly related to the {\it static} spin-spin correlator
\begin{equation}
 \chi_{ij}=\int_0^{\infty} d\tau \left < T_{\tau}\mathbf{S}_i(\tau)\cdot\mathbf{S}_j(0) \right >\;,\label{correlator}
\end{equation}
which can be derived by fusing the external legs $(1,1')$ and $(2,2')$ of $\Gamma^\Lambda(1',2';1,2)$. Exploiting translation invariance of the lattice and transforming the site variables $i_1$ and $i_2$ into $\mathbf{k}$-space yields the spin susceptibility $\chi^\Lambda(\mathbf{k})$ as a function of the RG scale $\Lambda$. The magnetic properties of the system can be deduced from the $\Lambda$ evolution of the susceptibility. In the case of magnetic long-range order with wave vector $\mathbf{k}$, the corresponding susceptibility grows as $\Lambda$ is decreased, until a peak or a kink indicates a magnetic instability breakdown of the RG flow (note that with a dense frequency grid and in the thermodynamic limit, i.e., without limiting the spatial extent of the two-particle vertex, these peaks would grow and eventually become divergences). Otherwise, a smooth flow that does not show signatures of an instability down to $\Lambda\rightarrow0$ indicates a magnetically disordered phase.

\subsection{Modifications for arbitrary spin length $S$} 
\label{sec:ls}

Our approach of generalizing the $S=1/2$ PFFRG method of the last section to arbitrary spin lengths $S$ amounts to considering multiple copies of spin-$1/2$ degrees of freedom on each site.\cite{affleck87_2,singh91,brinckmann04} In the first step we replace the spin operators $\mathbf{S}_i$ by a sum of $M$ spin flavors, i.e.
\be
\mathbf{S}_i\rightarrow\sum_{\kappa=1}^M \mathbf{S}_{i\kappa}\;,\label{sum}
\ee
where $\kappa$ denotes the new ``flavor'' index. Inserting this into the Hamitonian in Eq.~(\ref{hamiltonian}) we obtain
\be
\mathcal{H} = \sum_{(ij)}J_{ij}\left(\sum_{\kappa=1}^M \mathbf{S}_{i\kappa} \right)\cdot\left(\sum_{\kappa'=1}^M \mathbf{S}_{j\kappa'} \right)\label{ham_new}
\ee
showing that in this type of modified spin system, all flavors $\kappa$ on site $i$ interact with all flavors $\kappa'$ on site $j$ via the same coupling $J_{ij}$.

According to standard angular momentum addition rules, the sum of two arbitrary momenta $\mathbf{L}_1+\mathbf{L}_2$ defined in the product space $\mathds{V}^{L_1}\otimes\mathds{V}^{L_2}$ can be expressed in a basis such that $\mathbf{L}_1+\mathbf{L}_2$ decomposes into individual momenta with quantum numbers $|L_1-L_2|,|L_1-L_2|+1,\ldots,L_1+L_2$. One can therefore write the product space of two angular momenta as a direct sum,
\be
\mathds{V}^{L_1}\otimes\mathds{V}^{L_2}=\mathds{V}^{|L_1-L_2|}\oplus\mathds{V}^{|L_1-L_2|+1}\oplus\ldots\oplus\mathds{V}^{|L_1+L_2|}\;.
\ee
Successively adding up spin-$1/2$ angular momenta as in Eq.~(\ref{sum}), hence, yield series of the form
\begin{align}
&\mathds{V}^{1/2}\otimes\mathds{V}^{1/2}=\mathds{V}^{0}\oplus\mathds{V}^{1}\,,\notag\\
&\mathds{V}^{1/2}\otimes\mathds{V}^{1/2}\otimes\mathds{V}^{1/2}=\ldots\oplus\mathds{V}^{3/2}\;.\label{series}
\end{align}
It follows that the product space of $M$ spin-1/2 momenta on each lattice site can be written as a direct sum, where the highest angular momentum subspace $\mathds{V}^{M/2}$ appears exactly {\it once} while the other subspaces $\mathds{V}^{M/2-1},\mathds{V}^{M/2-2},\ldots$ might have larger multiplicities.

Applying the pseudo-fermionic representation in Eq.~(\ref{representation}) to set up a generalized spin-$S$ PFFRG scheme, the fermions acquire an extra flavor index,
\be
S_{i\kappa}^{\mu}=\frac{1}{2}\sum_{\alpha \beta}f_{i\alpha\kappa}^{\dagger}\sigma_{\alpha\beta}^{\mu}f_{i\beta\kappa}\,,\label{representation2}
\ee
where the operators $f_{i\alpha\kappa}$ fulfill the standard fermionic anti-commutation relation
\be
\{ f_{i\alpha\kappa},f^{\dagger}_{i'\alpha'\kappa'}\} = \delta_{i i'}\delta_{\alpha \alpha'} \delta_{\kappa \kappa'}\,.
\ee
Since the operators $f^{(\dagger)}_{i\uparrow\kappa}$, $f^{(\dagger)}_{i\downarrow\kappa}$ for a given site $i$ and flavor $\kappa$ generate angular momentum vector spaces $\mathds{V}^0\oplus\mathds{V}^0\oplus\mathds{V}^{1/2}$, summing up $M$ of these momenta now generates a direct sum containing all vector spaces $\mathds{V}^{0},\mathds{V}^{1/2},\ldots,\mathds{V}^{(M-1)/2},\mathds{V}^{M/2}$. Note that the multiplicities might be different as compared to the series in Eq.~(\ref{series}). The largest contribution $\mathds{V}^{M/2}$, however, still occurs exactly once.

Since we aim to use this approach to study spin models with a certain fixed spin $S$, we first need to find out in which of these subspaces the ground state of Eq.~(\ref{ham_new}) is realized (or whether it has contributions from different sectors). Given that the highest subspace $\mathds{V}^{M/2}$ yields the largest angular momentum eigenvalues, it is natural to assume that the ground state is constructed from states in $\mathds{V}^{M/2}$ on each site. We will show in Sec.~\ref{sc:local} that this is indeed the case by considering additional level repulsion terms $-(\sum_{\kappa=1}^M \mathbf{S}_{i\kappa})^2$ on the honeycomb lattice which further lower the energy of the highest angular momentum sector as compared to all other sectors. Based on these results we will conclude that the ground state of the modified Hamiltonian in Eq.~(\ref{ham_new}) with $M$ spin flavors is identical to the ground state of the model (\ref{hamiltonian}) with spin length $S=M/2$.

One important comment is in order. Instead of considering multiple copies of spin-1/2 degrees of freedom on each site, it might appear more straightforward to generalize the Pauli matrix representation $\sigma^\mu$ in Eq.~(\ref{representation}) to higher angular momenta, as described, e.g., in Ref.~\onlinecite{liu10}. In such a scheme, the implementation of a spin-$S$ degree of freedom requires the introduction of $2S+1$ fermions on each site with a pseudo fermion constraint fixing the particle number to either 1 or $2S$. In a situation where the free fermions do not disperse (i.e., they have zero band width) realizing an average occupation that is different from half filling poses a serious problem: Applying a finite chemical potential $\mu$ either depletes the system completely ($\mu>0$) or induces the maximal fermion occupation ($\mu<0$). In our scheme this problem is avoided since for each flavor $\kappa$ a spin-1/2 degree of freedom is realized at half filling which corresponds to a chemical potential $\mu=0$. 
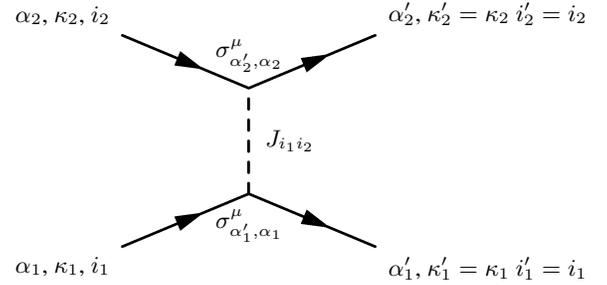
\begin{figure}
\begin{fmffile}{diagram2}
\bee
\begin{fmfgraph*}(120,80)
\fmfleft{e1,e2}
\fmfright{e3,e4}
\fmflabel{$\alpha$, $\kappa$}{e1}
\fmflabel{$\beta$, $\kappa$}{e2}
\fmflabel{$\alpha'$, $\kappa'$}{e3}
\fmflabel{$\beta'$, $\kappa'$}{e4}
\fmflabel{$j$}{v1}
\fmflabel{$i$}{v2}
\fmf{fermion}{e1,v1,e3}            %this two lines turns the diagram upside down
\fmf{fermion}{e2,v2,e4}
\fmf{dashes, label=$J_{ij}$}{v1,v2}
\end{fmfgraph*}
\eee
\end{fmffile}
\caption{Index structure of the bare two-particle vertex including the flavor index $\kappa$. The dashed line denotes the exchange couplings $J_{ij}$. Note that the site indices $i$ and flavor indices $\kappa$ do not change along fermion lines.}\label{fig:initialkappa}
\end{figure}

Setting up a diagrammatic theory with the new flavor indices $\kappa$ is now straightforward. The fundamental building blocks for Feynman diagrams are the bare propagator $G_0(1';1)$ and the bare interaction $\Gamma^\infty(1',2';1,2)$ (i.e., the two-particle vertex at $\Lambda\rightarrow\infty$). Instead of Eqs.~(\ref{g0}) and (\ref{initial}), they are now given by
\be
G_0(1';1)=\frac{1}{i\omega_1}\delta(\omega_1-\omega_{1'})\delta_{i_{1'} i_1}\delta_{\alpha_1 \alpha_{1'}}\delta_{\kappa_{1'}\kappa_1}\label{g0kappa}
\ee
and
\begin{align}
\Gamma^\infty(1',2';1,2)=&\frac{1}{4}J_{i_1 i_2}\sigma^\mu_{\alpha_{1'}\alpha_1}\sigma^\mu_{\alpha_{2'}\alpha_2}\delta_{i_{1'} i_1}\delta_{i_{2'} i_2}\delta_{\kappa_{1'}\kappa_1}\delta_{\kappa_{2'}\kappa_2}\notag\\
&\times\delta(\omega_1+\omega_2-\omega_{1'}-\omega_{2'})\notag\\
&-(\omega_1\leftrightarrow\omega_2,i_1\leftrightarrow i_2,\alpha_1\leftrightarrow\alpha_2,\kappa_1\leftrightarrow\kappa_2)\;.\label{initialkappa}
\end{align}
Here, the multi indices also include the $\kappa$ variables, i.e. ``$1=\{\omega_1,i_1,\alpha_1,\kappa_1\}$''. The index structure of the first term of Eq.~(\ref{initialkappa}) is illustrated in Fig.~\ref{fig:initialkappa}. Most importantly, Eqs.~(\ref{g0kappa}) and (\ref{initialkappa}) reveal that the index structures in $\kappa$ and $i$ are identical, indicating that the flavor index effectively behaves like a site variable. With this equivalence, the analog of Eq.~(\ref{parametrize}) is immediately given by
\begin{align}
\Gamma^\Lambda(1',2';1,2)= &\tilde\Gamma_{i_1i_2\,\kappa_1\kappa_2}^\Lambda(1',2';1,2)\delta_{i_{1'}i_1}\delta_{i_{2'}i_2}\delta_{\kappa_{1'}\kappa_1}\delta_{\kappa_{2'}\kappa_2}\notag\\
&-(\omega_1\leftrightarrow\omega_2,i_1\leftrightarrow i_2,\alpha_1\leftrightarrow\alpha_2,\kappa_1\leftrightarrow\kappa_2)\;.\label{parametrizekappa}
\end{align}
As noted earlier, the exchange couplings $J_{i_1 i_2}$ in Eq.~(\ref{initialkappa}) do not depend on the flavor variables such that there is no explicit $\kappa$ dependence in the scheme. Consequently, the couplings $J_{i_1 i_2}$ also remain independent of the flavor index on all levels of diagrammatic renormalizations yielding $\tilde\Gamma_{i_1i_2\,\kappa_1\kappa_2}^\Lambda(1',2';1,2)\equiv\tilde\Gamma_{i_1i_2}^\Lambda(1',2';1,2)$. With this, the modifications of the PFFRG scheme are rather simple: All terms in Eqs.~(\ref{FRG_sigma}) and (\ref{FRG_gamma}) that contain a site summation $\sum_j$ now also acquire a flavor sum $\sum_{\kappa=1}^M$ producing an extra factor $M$ in these terms. We therefore conclude that (given that the above assumption about the angular momentum subspace of the ground state is correct) a spin-$S$ generalization of the PFFRG only requires an additional prefactor $M=2S$ in the first term on the right-hand side of Eq.~(\ref{FRG_sigma}) and in the RPA channel of Eq.~(\ref{FRG_gamma}) (i.e. the second term on the right-hand side of this equation). This is a remarkable result as it shows that arbitrary spin lengths $S$ can be easily implemented in the PFFRG scheme without additional numerical efforts.
\begin{figure*}
\begin{minipage}{0.33\textwidth}
%\centering
\subfloat[]{
\includegraphics[scale=0.7,angle=0]{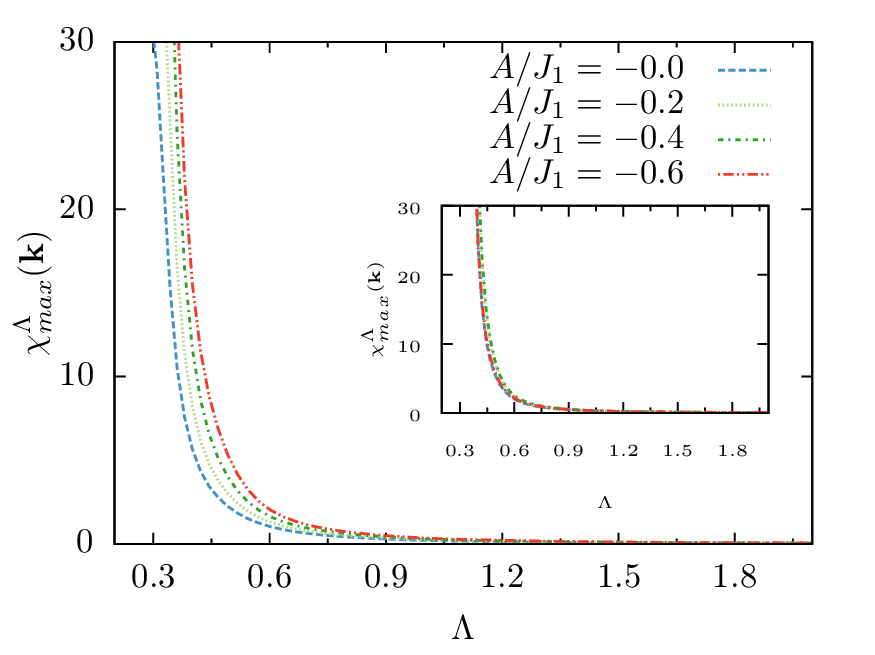}}
\end{minipage}
\begin{minipage}{0.33\textwidth}
\subfloat[]{
\includegraphics[scale=0.7,angle=0]{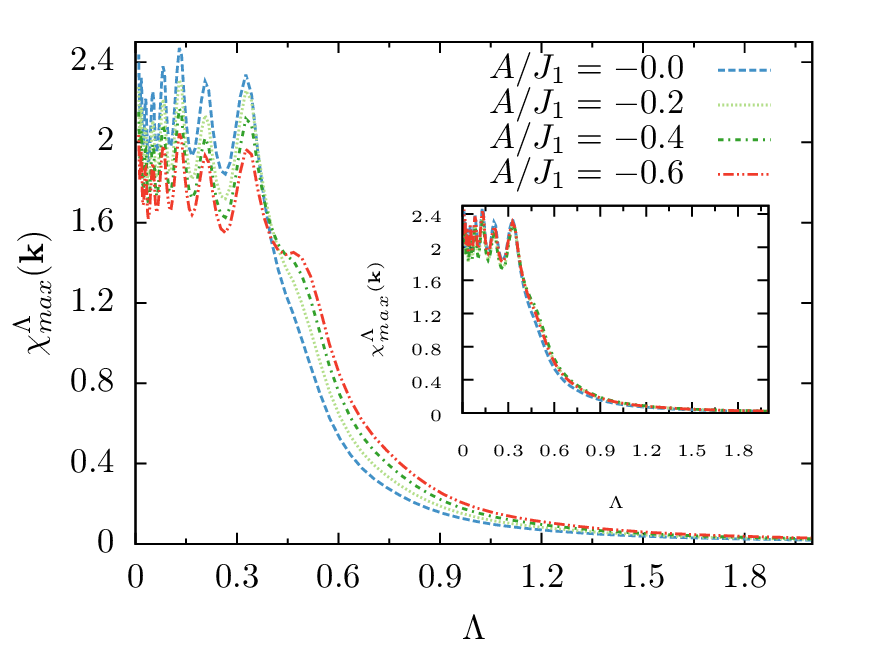}}
\end{minipage}
\begin{minipage}{0.33\textwidth}
%\centering
\subfloat[]{
\includegraphics[scale=0.7,angle=0]{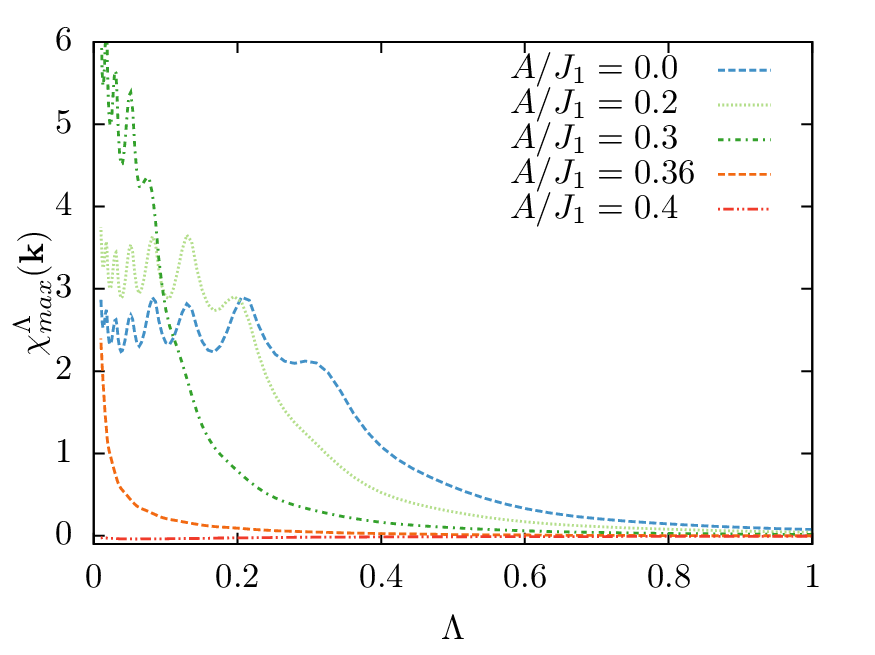}}
\end{minipage}
\caption{Flowing PFFRG susceptibility for the nearest neighbour honeycomb Heisenberg antiferromagnet with onsite level repulsion terms [see Eq.~(\ref{ham_a})]. In the main plots and in all subsequent figures, the susceptibility $\chi^\Lambda(\mathbf{k})$ (RG scale $\Lambda$) is given in units of $1/J_1$ ($J_1$). In the insets, energy scales are in units of $\sqrt{A^2+J_1^2}$ to compensate for energy renormalization effects in $\Lambda$ (see text for details). (a) Flow behavior of the susceptibility for the nearest neighbor model at $S=3/2$ and varying negative values of $A$ (due to almost diverging susceptibilities, the RG flow is not shown below the critical $\Lambda$ scale). Depicted is the maximal susceptibility component in $\mathbf{k}$ space which, here, corresponds to antiferromagnetic N\'eel order on the honeycomb lattice. (b) Same as in (a) but with $S=1/2$. (c) Susceptibility for $J_2/J_1 = 0.1$, $S=1/2$, and positive level repulsion terms $A\geq0$.\label{level_repulsion}}
 \end{figure*}

\subsection{Level repulsion terms}
\label{sc:local}

Above we have claimed that the ground state of the spin model in Eq.~(\ref{ham_new}) featuring $M$ copies of spin-1/2 degrees of freedom on each site is constructed from states in the highest angular momentum sector. A simple way to check this is by adding onsite spin terms to the Hamiltonian, yielding a spin model of the form
\be
\mathcal{H} = \sum_{(ij)}J_{ij}\left(\sum_{\kappa=1}^M \mathbf{S}_{i\kappa} \right)\cdot\left(\sum_{\kappa'=1}^M \mathbf{S}_{j\kappa'} \right)+A\sum_i \left(\sum_{\kappa=1}^M \mathbf{S}_{i\kappa} \right)^2\,.\label{ham_a}
\ee
The eigenvalues of the operator $(\sum_{\kappa=1}^M \mathbf{S}_{i\kappa})^2$ (with $\mathbf{S}_{i\kappa}$ expressed in terms of pseudo fermions) are given by $S(S+1)$ where the total angular momentum quantum number $S$ can take all values $0,1/2,\ldots,(M-1)/2, M/2$. When $A$ is chosen negative, all finite angular momentum sectors are shifted down in energy, with the largest energy reduction taking place in the highest sector with $S=M/2$. If our assumption is correct, further reducing the energy of the highest subspace with respect to the other ones should have no effects on our results. 

We tested this for the honeycomb Heisenberg antiferromagnet with nearest neighbor interactions $J_1>0$. A representative plot for $S=3/2$ is shown in Fig.~\ref{level_repulsion}(a). It can be seen that the susceptibility flow behavior remains qualitatively unchanged as $A$ is decreased from zero, except for an overall shift of the curves towards higher values of $\Lambda$. This behavior is expected since $A$ and $\Lambda$ both have the dimension of an energy. Increasing $|A|$ while keeping $J_1$ fixed increases the overall energy scale of the system such that the parameter $\Lambda$ becomes renormalized. To account for these effects, we repeated the calculations for rescaled values of $A$ and $J_1$. Phenomenologically, we find that for fixed $\sqrt{A^2+J_1^2}$ such artifacts are largely removed, yielding an approximate collapse of all curves, see inset in Fig.~\ref{level_repulsion}(a).

The investigation of level repulsion terms is particularly insightful for spin-1/2 systems since such models have been previously studied with PFFRG.\cite{reuther10,balz16,iqbal15,iqbal16,buessen16,hering16,reuther11,reuther11_2} In this case it can be tested whether unphysical spin-zero occupations such as singly and doubly occupied sites are indeed energetically suppressed in the ground state. As an example, we show in Fig.~\ref{level_repulsion}(b) the susceptibility flow behavior for the nearest neighbor honeycomb Heisenberg antiferromagnet for $S=1/2$. In analogy to the spin-3/2 case, the flow remains qualitatively unchanged and shifts in $\Lambda$ can again be compensated by keeping $\sqrt{A^2+J_1^2}$ constant [inset in Fig.~\ref{level_repulsion}(b)].

Additional calculations also confirm the absence of any qualitative changes in the RG flow for finite second neighbor interactions $J_2$ and varying $S$. In particular, phase boundaries between different magnetic phases or melting transitions into non-magnetic phases are never found to be affected by $A$. We therefore conclude that at least for the honeycomb Heisenberg model our above assumption is correct. Based on our experience with quantum spin systems on different lattices, we anticipate that also a wider class of spin models shares this property. For the spin-1/2 case, our analysis further shows that the average treatment of the pseudo fermion constraint in previous PFFRG studies was justified.

Another interesting situation arises when $A$ is positive. In this case, the energy levels in the highest angular momentum sector undergo the largest relative {\it increase}, until above a certain threshold of $A$, lower subspaces should become energetically preferred. The situation for $J_2/J_1=0.1$ and $S=1/2$ is depicted in Fig.~\ref{level_repulsion}(c), where the absolute value of $A$ is varied within similar ranges as in Figs.~\ref{level_repulsion}(a) and (b) but with a positive sign. Upon increasing $A$ we first observe a decrease of the critical $\Lambda$, followed by a sudden drop of the susceptibility at $A\approx0.35$, and almost vanishing responses above this value. We interpret this behavior as a consequence of promoting the unphysical zero or doubly occupied states. When $A$ is sufficiently large, the ground state resides entirely in the unphysical sector of the Hilbert space. Since these states carry $S=0$ and do not contribute to the magnetic susceptibility, the response is expected to vanish.

%====================================================================
%
\begin{figure*}
\begin{minipage}{0.4\textwidth}
%\centering
\hspace{-2cm}
\subfloat[]{
\includegraphics[scale=0.45,angle=0]{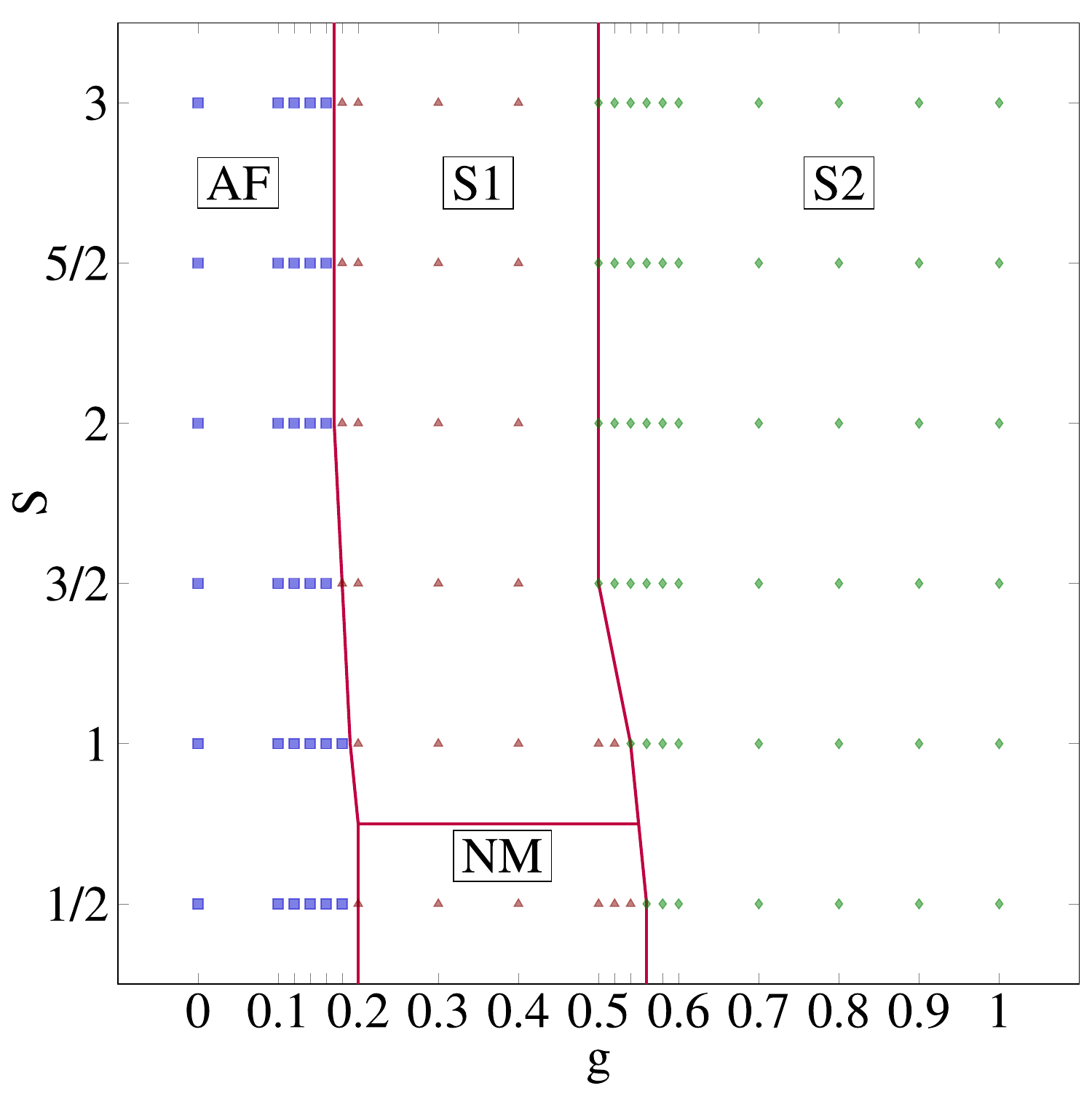}}
\end{minipage}
\begin{minipage}{0.4\textwidth}
%\centering
\subfloat[]{
\includegraphics[scale=0.93,angle=0]{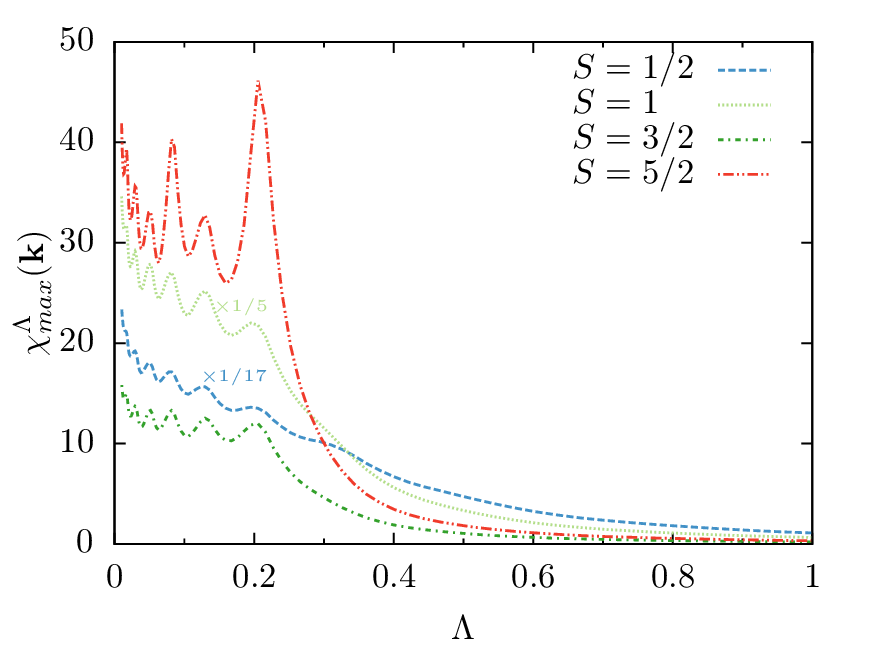}}
\end{minipage}
\caption{(a) Phase diagram in the $g$-$S$ plane obtained via PFFRG. We find a non-magnetic (NM) intermediate phase at $S=1/2$ and three magnetically ordered phases [antiferromagnetic (AF) state and two spiral phases S1, S2]. For the susceptibility profiles of the magnetic states in $\mathbf{k}$ space, see Figs.~\ref{profiles}(a)-(c) respectively. (b) $\Lambda$ flow of the maximal $\mathbf{k}$ space component of the susceptibility $\chi^\Lambda(\mathbf{k})$ for $g=0.3$ and increasing values of $S$. While the flow for $S=1/2$ does not show signatures of an instability, for $S\geq1$ we find a kink in the susceptibility at $\Lambda\approx0.2$.}\label{phasediag}
 \end{figure*}

\section{Antiferromagnetic $J_1$-$J_2$ Heisenberg model on the honeycomb lattice}\label{sec:j1j2}
 
\subsection{Phase diagram in the $J_2/J_1$-$S$ plane via PFFRG}\label{sec:phd}
We now apply the spin-$S$ generalization of the PFFRG method discussed in the last section to the antiferromagnetic $J_1$-$J_2$ Heisenberg model on the honeycomb lattice as illustrated in Fig.~\ref{lattice}. The Hamiltonian is given by
\be
H = J_1\sum_{\langle ij \rangle} \mathbf{S}_i\mathbf{S}_j + J_2\sum_{\langle \langle ij \rangle \rangle} \mathbf{S}_i\mathbf{S}_j\;,
\ee
where $\langle ij \rangle$ denotes a pair of nearest neighbor sites while $\langle\langle ij \rangle\rangle$ indicates second neighbor sites. The corresponding exchange couplings are $J_1>0$ and $J_2\geq 0$, respectively. The ratio of the two couplings is denoted by $g=J_2/J_1$.

Numerically solving the PFFRG equations for varying parameters in the $g$-$S$ plane we obtain the phase diagram shown in Fig.~\ref{phasediag}(a). For $S=1/2$ we reproduce the phases that have previously been found within PFFRG, see Ref.~\onlinecite{reuther11_2}: An extended non-magnetic phase at $g\approx0.2\ldots0.6$ is framed by an antiferromagnetic phase at $0\leq g\lesssim 0.2$ and an incommensurate spiral phase at $g\gtrsim0.6$. When $S$ is increased, the phase diagram changes drastically. Already at $S=1$, the non-magnetic phase is completely eaten up by spiral magnetic long-range order. This leads, in total, to three magnetically ordered phases at $S=1$: An antiferromagnetically ordered regime at $0\leq g\lesssim 0.19$ and two spiral phases S1, S2 at $0.19\lesssim g\lesssim 0.53$ and $g\gtrsim0.53$, respectively, whose nature will be discussed in more detail below. While this sequence of phases persists for larger values of $S$, the locations of the two phase transitions shift towards the classical values 1/6 and 0.5, see Fig.~\ref{phasediag}(a).

To demonstrate the onset of magnetic long-range order for all spin lengths $S\geq 1$ we show in Fig.~\ref{phasediag}(b) the PFFRG flow of the susceptibility for the highly frustrated case $g=0.3$ and varying values of $S$. While at $S=1/2$ we do not observe an instability feature as $\Lambda$ is decreased, hinting at a magnetically disordered phase, for all values $S\geq 1$ we find pronounced cusps at $\Lambda\approx0.2$ associated with the onset of a numerically uncontrolled, oscillating flow behavior. Within PFFRG, such features indicate that in the thermodynamic limit the system would run into a magnetic instability. The point in $\mathbf{k}$ space at which this breakdown occurs further determines the type of magnetic order. With increasing $S$ the susceptibility grows and the cusp becomes more pronounced, signaling an increase of the ordered magnetic moment.
\begin{figure*}
\begin{minipage}{0.33\textwidth}
%\centering
\includegraphics[scale=0.7,width=62mm, height=50mm, angle=0]{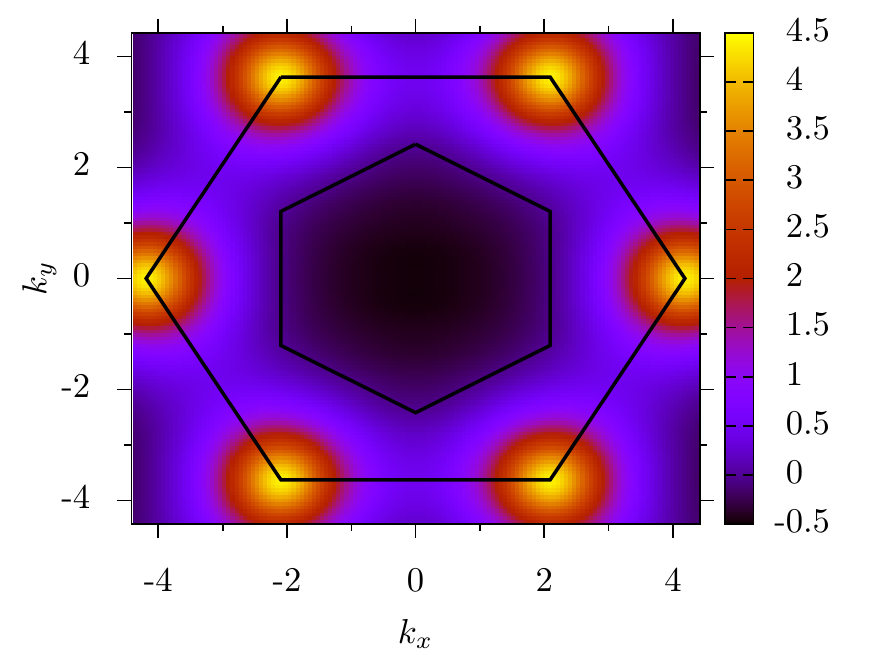}
\end{minipage}
\begin{minipage}{0.33\textwidth}
%\centering
\includegraphics[scale=0.7,width=62mm, height=50mm, angle=0]{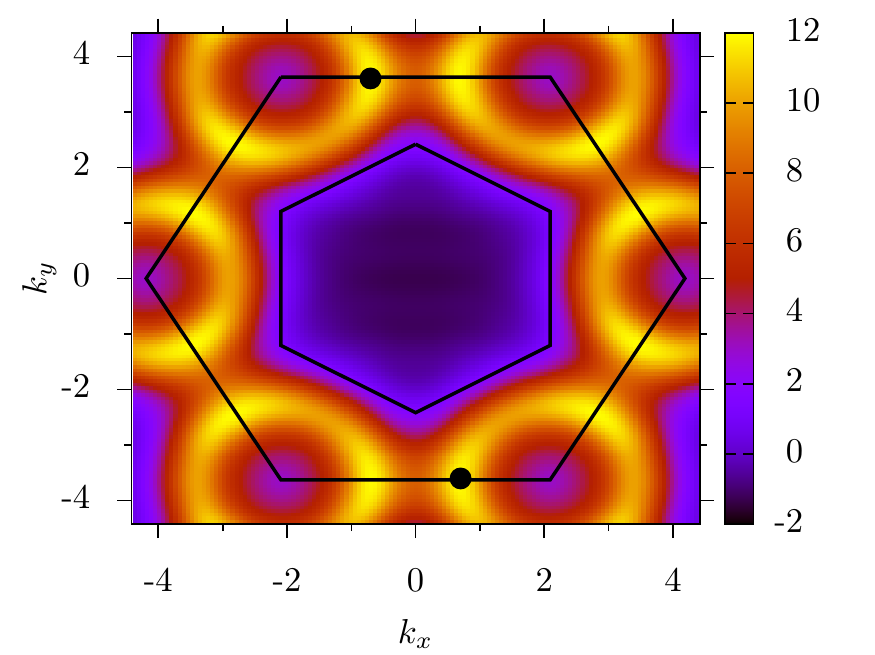}
\end{minipage}
\begin{minipage}{0.33\textwidth}
%\centering
\includegraphics[scale=0.7,width=62mm, height=50mm, angle=0]{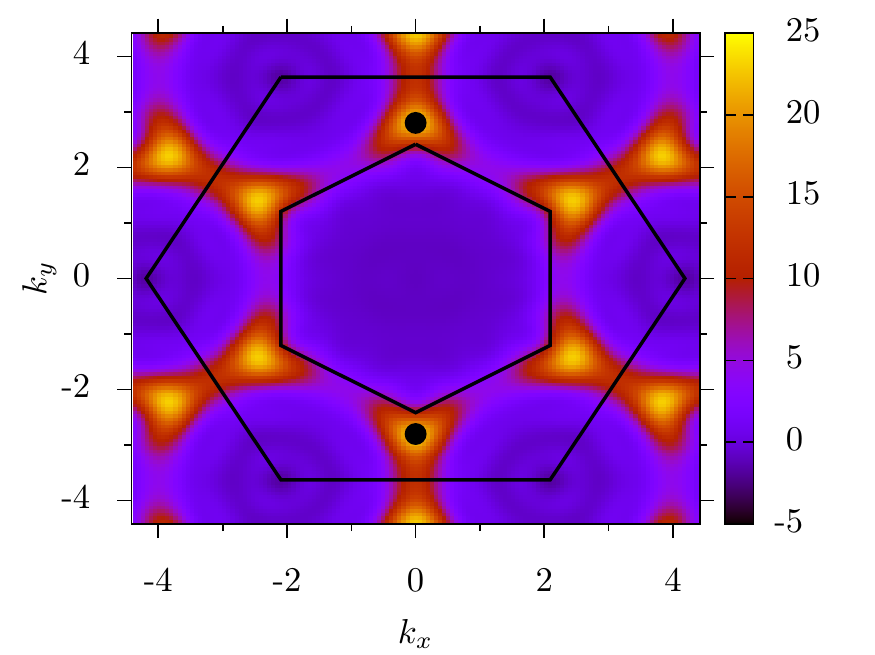}
\end{minipage}
\end{figure*}

\begin{figure*}
\begin{minipage}{0.33\textwidth}
%\centering
\subfloat[]{
\includegraphics[scale=0.8,angle=0]{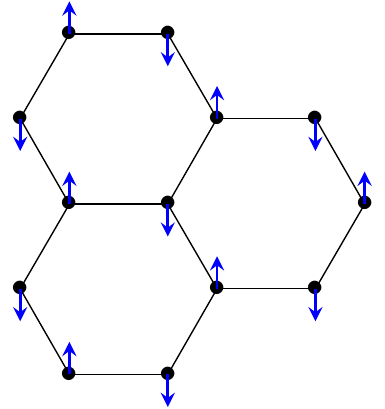}}
\end{minipage}
\begin{minipage}{0.33\textwidth}
%\centering
\subfloat[]{
\includegraphics[scale=0.8,angle=0]{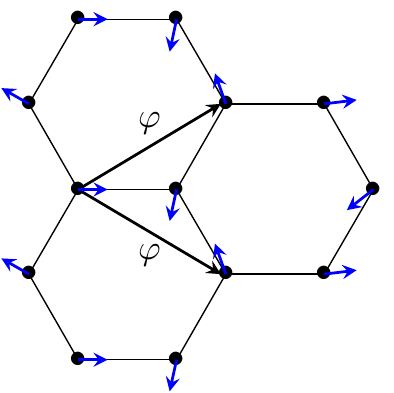}}
\end{minipage}
\begin{minipage}{0.33\textwidth}
%\centering
\subfloat[]{
\includegraphics[scale=0.8,angle=0]{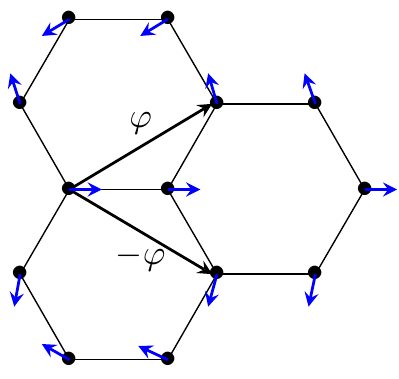}}
\end{minipage}
\caption{Upper panel: Susceptibility $\chi^\Lambda(\mathbf{k})$ in reciprocal space for the three magnetically ordered phases at $S=3/2$. (a) antiferromagnetic state at $g=0$,  (b) S1 spiral at $g=0.3$, and (c) S2 spiral at $g=0.9$. All plots correspond to $\Lambda$ values right above the instability feature during the RG flow. Outer (inner) hexagons indicate the boundaries of the extended (first) Brillouin zone. Lower panel: Below each susceptibility profile we depict the corresponding real space spin patterns which yield magnetic Bragg peaks in $\mathbf{k}$ space at the marked positions (black dots). Arrows illustrate the unit vectors of the honeycomb lattice and indicate the pitch angles of the spiral state along these directions.}\label{profiles}
\end{figure*}

To study in more detail the types of magnetic orders detected in the system, we plot in Fig.~\ref{profiles} the $\mathbf{k}$ space resolved susceptibilities at $S=3/2$ within the three ordered phases, along with real space illustrations of the spin patterns. In the antiferromagnetic phase [Fig.~\ref{profiles}(a)] sharp magnetic Bragg peaks are located at the corners of the extended Brillouin zone. As $g$ is increased, the system first establishes planar incommensurate spiral order of S1 type which is characterized by magnetic wave vectors residing at the edges of the extended Brillouin zone, as shown in Fig.~\ref{profiles}(b). The susceptibility profile in this phase exhibits pronounced ring-like features. Along these rings the magnetic wave vectors only correspond to small maxima at the Brillouin zone edges, in agreement with the quantum selection described in Ref.~\onlinecite{mulder10}. These correlations already resemble the continuous set of degenerate ground states expected in the classical limit. To depict this spin state in real space [Fig.~\ref{profiles}(b)] we construct a planar spiral which -- upon Fourier transformation -- yields a dominant Bragg peak in $\mathbf{k}$ space at exactly the position of the maximum of the PFFRG susceptibility. As a characteristic feature of this state, the spiral pitch angles along the lattice vectors indicated in Fig.~\ref{profiles}(b) are identical. Further increasing $g$ the system enters the S2 spiral phase, which shows magnetic Bragg peaks at the $k_x=0$ line (or symmetry related positions), see Fig.~\ref{profiles}(c). These peaks correspond to a planar spiral with pitch angles of opposite signs but same absolute value. Furthermore, along one of the three nearest neighbor directions, pairs of spins are in parallel orientation.

\begin{figure}
\includegraphics[scale=0.2,angle=0]{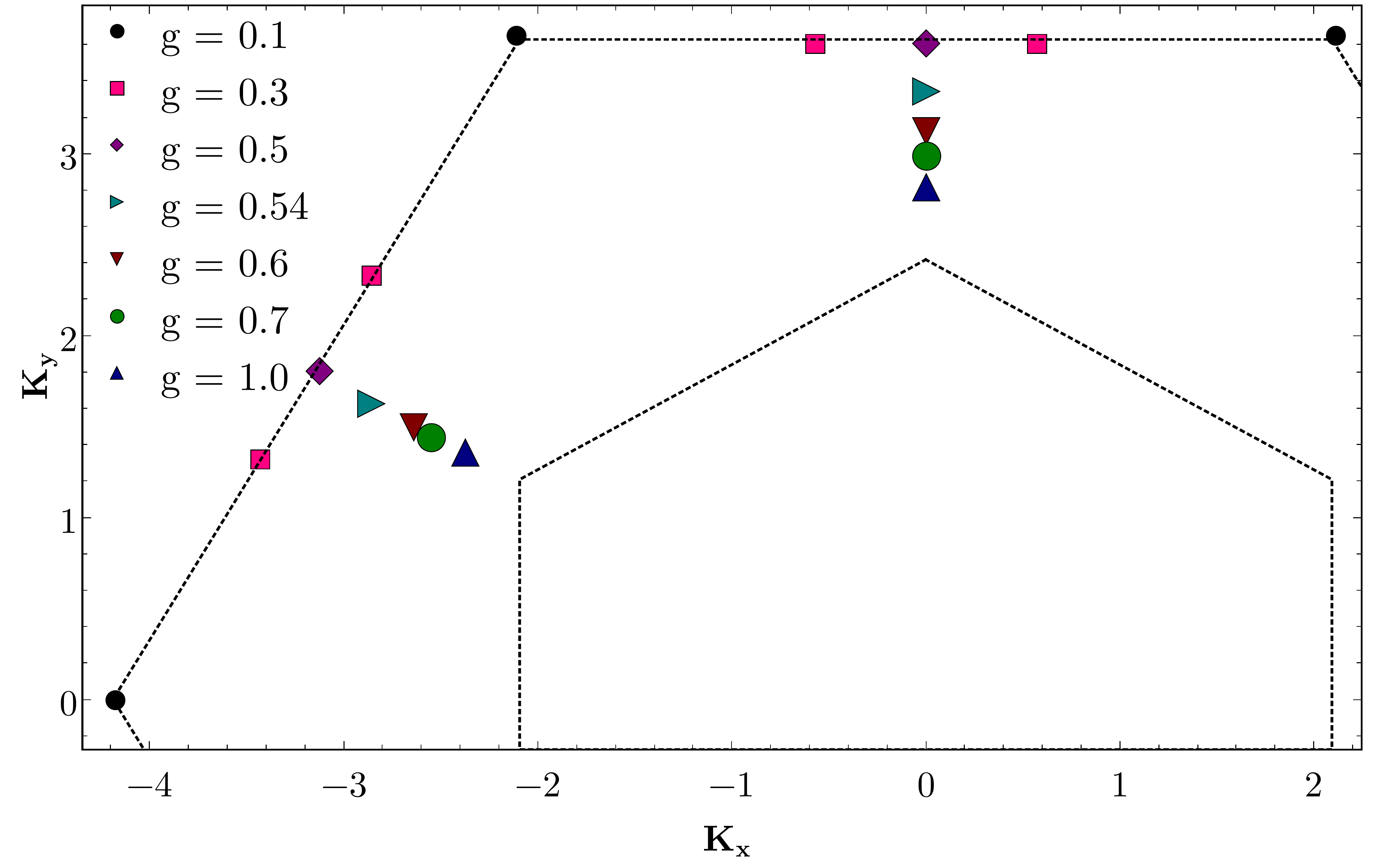}
\caption{Position of the magnetic wave vectors in reciprocal space at $S=3/2$ for increasing values of $g$: The antiferromagnetic state is characterized by susceptibility peaks at the corners of the extended Brillouin zone. In the S1 and S2 spiral phases the maxima move along the Brillouin zone edges and along a radial direction, respectively.}
\label{move}
\end{figure}

The overall migration of the magnetic wave vectors in $\mathbf{k}$ space upon increasing $g$ is illustrated in Fig.~\ref{move} for $S=3/2$. In the antiferromagnetic phase the magnetic Bragg peaks remain at the corner position of the extended Brillouin zone and start moving along the Brillouin zone boundary as the system enters the S1 phase. At the transition between the S1 and S2 spirals, the peaks reside exactly at the midpoints of the edges. Further increasing $g$, they move towards the center and reach the corners of the first Brillouin zone in the limit $g\rightarrow\infty$. This position corresponds to 120$^\circ$ N\'eel order on the triangular lattice which is realized when the two sublattices of the honeycomb lattice decouple.  

\subsection{Classical large $S$ limit} 
\label{sec:largeS}
\subsubsection{RPA solution}
To shed more light on the spin-$S$ generalization of the PFFRG method and the approximations associated with it, we now consider the classical limit $S\rightarrow\infty$ where the flow equations can be solved analytically. Starting from the PFFRG equations (\ref{FRG_sigma}) and (\ref{FRG_gamma}), we have argued that an arbitrary spin length $S$ can be implemented by multiplying all interaction channels containing an internal closed fermion loop with a factor $M=2S$. Strengthening these terms relative to all other channels, the classical  limit is effectively described by RG equations in which only loop diagrams contribute, leading to 
\be
\frac{d}{d\Lambda}\Sigma_{i_1}^{\Lambda}\left(\omega_1\right)=-\frac{1}{2\pi}\sum\limits_{2}\sum_j\tilde\Gamma_{i_1j}^{\Lambda}(1,2;1,2)S_j^{\Lambda}(\omega_2)\;,\label{FRG_sigma2}
\ee
\begin{align}
&\frac{d}{d\Lambda}\tilde\Gamma_{i_1i_2}^{\Lambda}(1',2';1,2)\notag\\
&=-\frac{1}{2\pi}\sum_{3\,4}\sum_j\tilde\Gamma_{i_1j}^{\Lambda}(1',4;1,3)\tilde\Gamma_{ji_2}^{\Lambda}(3,2';4,2)P^\Lambda_{jj}(\omega_3,\omega_4)\,.\label{FRG_gamma2}
\end{align}
Here, we have omitted the prefactors $M$ to avoid diverging terms at $S\rightarrow\infty$. Due to the special spin-index structure of Eq.~(\ref{FRG_gamma2}), the property $\tilde\Gamma^\infty_{i_1 i_2}(1',2';1,2)\propto\sigma^\mu_{\alpha_{1'}\alpha_1}\sigma^\mu_{\alpha_{2'}\alpha_2}$ of the initial conditions [see Eq.~(\ref{initial2})] is retained during the entire RG flow (this is in contrast to the full PFFRG scheme where also density terms $\propto\delta_{\alpha_{1'}\alpha_1}\delta_{\alpha_{2'}\alpha_2}$ are generated). Hence, Eq.~(\ref{FRG_sigma2}) contains a vanishing spin sum $\sum_{\alpha_2}\sigma^\mu_{\alpha_{1}\alpha_1}\sigma^\mu_{\alpha_{2}\alpha_2}=0$ such that the self energy and the Katanin contribution [see Eq.~(\ref{eq:Katanin})] remain identically zero. Examining the frequency arguments in Eq.~(\ref{FRG_gamma2}) one further finds that the static component $\omega_{1'}=\omega_{2'}=\omega_{1}=\omega_{2}=0$ of the two-particle vertex completely decouples from all other components which allows us to perform the frequency integration analytically. This yields a flow equation of the form
\be
\frac{d}{d\Lambda}\tilde{\Gamma}^\Lambda_{i_1 i_2}=\frac{2}{\pi\Lambda^2}\sum_j \tilde{\Gamma}^\Lambda_{i_1 j}\tilde{\Gamma}^\Lambda_{j i_2}\label{FRG_gamma3}
\ee
where $\tilde{\Gamma}^\Lambda_{i_1 i_2}$ (without arguments ``1'', ``2'', $\ldots$) parametrizes the static two-particle vertex component via
\be
\tilde\Gamma^\Lambda_{i_1 i_2}(1',2';1,2)\big|_{\omega_{1'}=\omega_{2'}=\omega_{1}=\omega_{2}=0}=\tilde{\Gamma}^\Lambda_{i_1 i_2} \sigma^\mu_{\alpha_{1'}\alpha_1}\sigma^\mu_{\alpha_{2'}\alpha_2}\,.
\ee
This vertex is initially given by $\tilde{\Gamma}^\infty_{i_1 i_2}=\frac{1}{4}J_{i_1 i_2}$.
\begin{figure}
\includegraphics[width=0.99\linewidth]{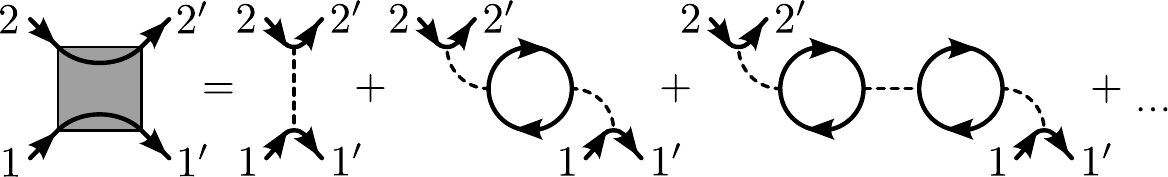}
\caption{Fermionic two-particle vertex in RPA approximation: Dashed lines denote bare exchange couplings $J_{i_1 i_2}$ and arrows illustrate free fermion propagators.}
\label{rpa}
\end{figure}

To simplify the remaining spatial dependence of Eq.~(\ref{FRG_gamma3}) we Fourier-transform $\tilde{\Gamma}^\Lambda_{i_1 i_2}$ using
\begin{equation}
\tilde\Gamma^\Lambda_{a(i)b(j)}(\mathbf{k})=\sum_{\Delta \mathbf{R}=\mathbf{R}_i-\mathbf{R}_j}e^{-i\mathbf{k}(\mathbf{R}_i-\mathbf{R}_j)}\tilde\Gamma^\Lambda_{ij}\;.\label{fourier}
\end{equation}
Here, $a(i)=1,2$ denotes a function that returns the sublattice index of site $i$ on the honeycomb lattice [$b(j)$ is defined in the same way] and $\mathbf{R}_i$ is the position of the two-site unit cell that contains site $i$. Since the Fourier-transform is only performed with respect to the unit-cell coordinates without involving the sublattice positions, different $\mathbf{k}$ components in the flow equations decouple, yielding
\be
\tilde\Gamma^\Lambda(\mathbf{k})=\frac{2}{\pi\Lambda^2}\left[\tilde\Gamma^\Lambda(\mathbf{k})\right]^2\,.\label{FRG_gamma4}
\ee
In this equation the vertex $\tilde\Gamma^\Lambda(\mathbf{k})$ is understood as a $2\times2$ matrix in the sublattice indices and the square on the right-hand side is a standard matrix product. The analytical solution of Eq.~(\ref{FRG_gamma4}) is given by
\be
\tilde\Gamma^{\Lambda}(\mathbf{k}) = \left[\frac{2}{\pi\Lambda}\mathds{1}_{2\times2}+\left(\tilde\Gamma^{\infty}(\mathbf{k})\right)^{-1}\right]^{-1}\,,
\label{analyt}
\ee
where $\tilde\Gamma^{\infty}(\mathbf{k})$ is the Fourier-transform of the bare exchange couplings $\frac{1}{4}J_{ij}$ using Eq.~(\ref{fourier}) and $\mathds{1}_{2\times2}$ denotes the two dimensional identity matrix. This equation has the form of an RPA solution and it can indeed be shown that the result is identical to an RPA summation in the pseudo fermions, as illustrated in Fig.~\ref{rpa}. The equivalence of the PFFRG and the pseudo fermion RPA in the limit $S\rightarrow\infty$ can also be understood from a pure diagrammatic picture: For each given order in the exchange couplings $J$, the RPA terms are those diagrams with the maximal number of closed fermion loops (in Fig.~\ref{rpa}, the $n$-th term on the right hand side is of $n$-th order in $J$ and contains $n-1$ loops). Since each loop contributes a factor $M$, the RPA diagrams are naturally singled out at $S\rightarrow\infty$.

The key outcome of Eq.~(\ref{analyt}) is the wave vector $\mathbf{k}_\text{RPA}$ at which the two-particle vertex diverges first as $\Lambda$ is decreased, determining the type of magnetic order the system develops in the classical limit. (Note that in contrast to the full PFFRG scheme where instabilities are typically signaled by kinks during the RG flow, here they appear as real divergencies.) Interestingly, Eq.~(\ref{analyt}) implies a simple scheme for finding $\mathbf{k}_\text{RPA}$, based on a minimization of the eigenvalues of the initial interaction matrix $\tilde\Gamma^\infty(\mathbf{k})$. We will call these eigenvalues $\lambda_m(\mathbf{k})$ below. Since the following arguments also prove the equivalence of the RPA and the Luttinger-Tisza method (as explained below), we generalize the discussion to arbitrary lattices with $n$ sites per unit cell. All relations in Eqs.~(\ref{fourier})-(\ref{analyt}) then become $n\times n$ matrix equations.

We first denote the eigenvalues of the matrix $\frac{2}{\pi\Lambda}\mathds{1}_{n\times n}+[\tilde\Gamma^{\infty}(\mathbf{k})]^{-1}$ by $\lambda'_m(\mathbf{k})$ and the eigenvalues of $[\tilde\Gamma^{\infty}(\mathbf{k})]^{-1}$ are given by $1/\lambda_m(\mathbf{k})$ (with $m=1,\ldots,n$). The term $\frac{2}{\pi\Lambda}\mathds{1}_{n\times n}$ only leads to an overall shift of these eigenvalues such that
\be
\lambda'_m(\mathbf{k})=\frac{2}{\pi\Lambda}+\frac{1}{\lambda_m(\mathbf{k})}\;.\label{diverging_gamma1}
\ee
According to Eq.~(\ref{analyt}) the two-particle vertex $\tilde\Gamma^\Lambda(\mathbf{k})$ diverges when the matrix $\frac{2}{\pi\Lambda}\mathds{1}_{n\times n}+[\tilde\Gamma^{\infty}(\mathbf{k})]^{-1}$ has a vanishing eigenvalue $\lambda'_m(\mathbf{k})$ at some wave vector $\mathbf{k}$. Setting $\lambda'_m(\mathbf{k})=0$ in Eq.~(\ref{diverging_gamma1}) the condition for this becomes
\be
\Lambda=-\frac{2}{\pi}\lambda_m(\mathbf{k})\;.\label{diverging_gamma2}
\ee
It follows that each negative eigenvalue $\lambda_m(\mathbf{k})<0$ can cause a diverging vertex $\tilde\Gamma^{\Lambda}(\mathbf{k})$ when Eq.~(\ref{diverging_gamma2}) is fulfilled. This condition also indicates that as $\Lambda$ is decreased from infinity, the {\it first} divergence occurs when the {\it smallest} (negative) eigenvalue $\lambda_m(\mathbf{k})$ satisfies Eq.~(\ref{diverging_gamma2}) (here, the term ``{\it smallest}'' refers to a minimization with respect to $\mathbf{k}$ and $m$). This proves that the classical magnetic order found within an RPA scheme occurs at the wave vector $\mathbf{k}_{\text{RPA}}$ that minimizes the eigenvalues of $\tilde\Gamma^{\infty}(\mathbf{k})$.
\begin{figure}
\centering
\includegraphics[scale=0.19,angle=0]{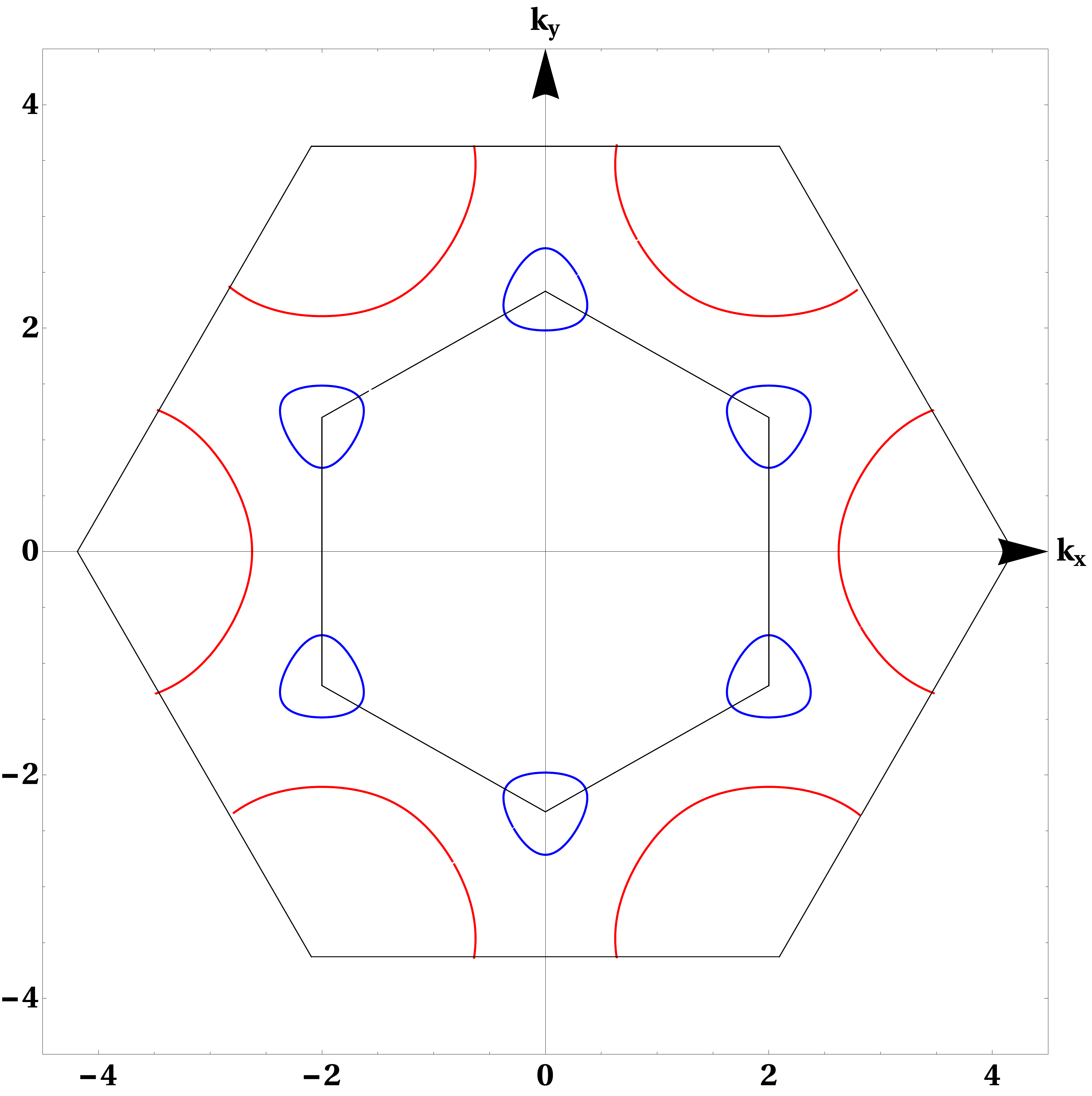}
\caption{Degenerate spiral magnetic wave vectors of the classical antiferromagnetic $J_1$-$J_2$ honeycomb Heisenberg model in reciprocal space. At $1/6<g<1/2$ the degenerate states form contours around the corners of the extended Brillouin zone, see red ring for $g=0.3$. For $g>0.5$ the contours are around the corners of the first Brillouin zone, see blue ring for $g=0.9$.}
\label{classlim}
\end{figure}

Expanded in terms of Pauli and identity matrices the initial two-particle vertex $\tilde\Gamma^\infty(\mathbf{k})$ for the antiferromagnetic $J_1$-$J_2$ honeycomb Heisenberg model is given by
\be
\tilde\Gamma^\infty(\mathbf{k})=\gamma^0(\mathbf{k})\mathds{1}_{2\times2}+\gamma^x(\mathbf{k})\sigma^x+\gamma^y(\mathbf{k})\sigma^y\;,\label{initial_fourier}
\ee
with
\begin{align}
\gamma^0(\mathbf{k})&=\frac{J_2}{2}\left(\cos k_+ +\cos k_- +\cos \sqrt{3}k_y\right)\;,\notag\\
\gamma^x(\mathbf{k})&=\frac{J_1}{4}\left(1+\cos k_+ +\cos k_-\right)\;,\notag\\
\gamma^y(\mathbf{k})&=\frac{J_1}{4}\left(\sin k_+ +\sin k_-\right)\;,
\end{align}
and
\be
k_\pm=\frac{3k_x}{2}\pm\frac{\sqrt{3}k_y}{2}\;.
\ee
Here, the nearest neighbor lattice distance is set to one and the sublattice structure follows the convention of Fig.~\ref{lattice}. We calculated the wave vectors $\mathbf{k}_\text{RPA}$ for arbitrary $g$ and compared the results with Ref.~\onlinecite{mulder10}, where the exact phase diagram is determined via a direct minimization of the classical energy. Throughout the phase diagram we find perfect agreement of the two approaches demonstrating that for the $J_1$-$J_2$ honeycomb Heisenberg model the spin-$S$ generalization of the PFFRG approach becomes exact. As discussed below, however, this exactness is not guaranteed for all classical spin models but depends on details of the magnetic states. We will argue that for the honeycomb Heisenberg model, the correctness is rooted in the sublattice symmetry of the system. 

To summarize the classical phase diagram, for small $g$ the system shows antiferromagnetic order which remains stable up to $g=1/6$. At $1/6<g<1/2$ one finds contours of degenerate classically ordered states forming rings in $\mathbf{k}$ space around the antiferromagnetic order position. With increasing $g$ the rings become larger, merge at $g=0.5$ and then form new rings around the corners of the first Brillouin zone, see Fig.~\ref{classlim}.

\subsubsection{Equivalence to the Luttinger-Tisza method}
The pseudo fermion RPA scheme can be put into a broader context when realizing that for arbitrary classical two-body spin models this approach is identical to the LT approximation.\cite{luttinger46,luttinger51} For complicated non-Bravais lattices and/or anisotropic interactions, even classical spin models may pose serious theoretical problems. In such situations, the LT method provides a simple framework to construct approximate classical ground states.\cite{lapa12,kimchi14,sklan13,nishimoto16} Instead of minimizing the classical energy under the hard constraint $|\mathbf{S}_i|^2=S$, normalizing the spin length on each site $i$ separately, the minimization is done subject to a weak constraint of the form
\be
\sum_i |\mathbf{S}_i|^2 = SN\,,
\ee
where $N$ is the total number of lattice sites. It can be shown\cite{lapa12} that with this condition the problem reduces to the same minimization of eigenvalues $\lambda_m(\mathbf{k})$ of $\tilde\Gamma^\infty(\mathbf{k})$ that yields the RPA solution. In the context of the LT method, the wave vector $\mathbf{k}_\text{LT}\equiv\mathbf{k}_\text{RPA}$ that minimizes $\lambda_m(\mathbf{k})$ is referred to as ``optimal'' LT eigenmode. If there exists a degenerate set of these modes (as is the case for the $J_1$-$J_2$ honeycomb Heisenberg model), classical ground states can be constructed by linear superpositions of the corresponding plane waves. The key question is whether it is possible to construct a state which (apart from the weak constraint) also fulfills the strong constraint. If this is the case, such a configuration represents the exact solution of the classical problem. At least for Bravais lattices it can be proven that a normalized spin state can always be formed with the eigenmodes $\mathbf{k}_\text{LT}$. The $J_1$-$J_2$ honeycomb model discussed here is an example where the LT method even works for a non-Bravais lattice. On more complicated lattices such as the 3d pyrochlore lattice, however, one finds that the modes $\mathbf{k}_\text{LT}$ are not sufficient to obtain a normalized spin state.\cite{lapa12,sklan13} To also satisfy the strong constraint, finite admixtures from ``suboptimal'' modes are required which do not correspond to the absolute minimum of $\lambda_m(\mathbf{k})$. While in such situations the LT method is no longer exact, the wave vectors $\mathbf{k}_\text{LT}$ still allow to construct phase diagrams of classical spin models which typically closely resemble the exact ones. The LT approach can therefore be used as a simple scheme to determine (at least) the dominant types of classical ordering.

The reason why the LT (and RPA) method works for the $J_1$-$J_2$ honeycomb Heisenberg model can be traced back to the equivalence of the two sublattices. With this property $\tilde\Gamma^\infty_{11}(\mathbf{k})=\tilde\Gamma^\infty_{22}(\mathbf{k})$ and Eq.~(\ref{initial_fourier}) has no contribution from $\sigma^z$. It follows that all eigenvectors $u^m(\mathbf{k})$ of $\tilde\Gamma^\infty(\mathbf{k})$ have sublattice components with equal norm, i.e., $|u_1^m(\mathbf{k})|^2=|u_2^m(\mathbf{k})|^2$. This allows one to superimpose plane wave modes with $\mathbf{k}_\text{RPA}$ and $-\mathbf{k}_\text{RPA}$ yielding states with normalized spins on both sublattices, hence fulfilling the strong constraint.

One puzzling aspect of the pseudo fermion RPA finally deserves to be clarified. Above we have argued that among all possible pseudo-fermion Feynman diagrams, the RPA terms are the only ones that survive in the classical limit. This is because the RPA diagrams maximize the number of closed fermion loops in each order in the exchange couplings. Since these diagrams are completely summed up within the PFFRG, one would expect that the RPA (and therefore also the LT method) is always exact at $S\rightarrow\infty$. It might therefore appear contradicting that there are also spin models where the LT method does not provide the correct classical state. This can be resolved by noting that FRG schemes generally only yield physical results in the cutoff-free limit $\Lambda=0$. The instabilities discussed here, however, occur at a finite critical $\Lambda=\Lambda_\text{c}$ such that the RG flow has to be stopped before the physical limit $\Lambda=0$ is reached. Therefore, any result obtained at a finite RG scale may still be subject to errors. In other words, the RPA scheme can be considered as classically exact above the instability in the sense that at each $\Lambda>\Lambda_\text{c}$ the correct and full amount of classical diagrams is included. The problem, however, is to reach $\Lambda=0$ within this approach. There are proposals to track the FRG flow into symmetry broken phases, which has been demonstrated for superconductivity in a BCS model.\cite{salmhofer04} For the PFFRG this would mean that time-reversal broken fields with $\Lambda$ dependent wave vectors have to be included explicitly, which represents an enormous complication of the method. In such type of generalization the self energy would no longer vanish at $S\rightarrow\infty$ but the loop term in Eq.~(\ref{FRG_sigma2}) would contribute. If such a scheme could be implemented, these terms would remove possible errors from missing suboptimal LT eigenmodes leading to exact classical results at $\Lambda=0$.

%====================================================================

\section{Conclusion and discussion}\label{sec:conclusion}

In this work, we have developed a general framework that allows one to study spin systems of arbitrary spin length $S$ within the PFFRG approach. Systems with $S>1/2$ are implemented by considering $M$ copies of spin-1/2 degrees of freedom on each lattice site. It is demonstrated that even without onsite level-repulsion terms, spin systems tend to realize the largest possible local spin magnitude $S=M/2$ in the ground state such that no further projection is necessary to fix the spin length. This also has important consequences for the $S=1/2$ case as it shows that single pseudo fermion occupancy is automatically satisfied in the ground state, justifying the average treatment of the particle constraint in previous PFFRG studies.

We have applied this method to the antiferromagnetic $J_1$-$J_2$ honeycomb Heisenberg model, mapping out the magnetic phase diagram as a function of $g=J_2/J_1$ and $S$. While for $S=1/2$ the frustrating effect of the $J_2$ interaction is strong enough to clearly indicate a magnetically disordered phase at $0.2\lesssim g \lesssim 0.6$, we find that for larger spins the phase diagram quickly resembles the classical one. In particular, already at $S=1$ the PFFRG does not detect any non-magnetic phases but instead shows clear signatures of two spiral magnetic phases at $g\gtrsim0.2$ and an antiferromagnetic phase at $g\lesssim0.2$. A characteristic feature of the momentum resolved spin susceptibility in the spiral phases are rings of strong response. Increasing $S$ the signal becomes more evenly distributed along the ridges of these rings and residual discrete maxima disappear. Our results can be benchmarked at $S\rightarrow\infty$ where the RG equations allow for an analytical solution. In this limit we exactly reproduce the known phase diagram of the classical system. Particularly, we show that spiral instabilities occur simultaneously for all wave vectors along rings in $\mathbf{k}$ space, in agreement with the degeneracy of classical states. More generally, we prove that for $S\rightarrow\infty$ the PFFRG method becomes identical to the LT approach.  

Comparing our results with other methods, the $S=1/2$ case has already been discussed in an earlier PFFRG work.\cite{reuther11_2} The existence of a non-magnetic intermediate phase is supported by the vast majority of previous studies.\cite{mulder10,oitmaa11,albuquerque11,fouet01,mosadeq11,yu14,zhu13,ganesh13,gong13,zhang13,bishop13,li12,mezzacapo12,bishop12} Whereas the precise extent and nature of this regime are not yet completely settled, the general tendency for the formation of staggered dimer order near the upper phase boundary is in agreement with many other works.\cite{mosadeq11,zhu13,ganesh13,gong13,zhang13,bishop13} Comparing results in the opposite limit $S\rightarrow\infty$, the continuous set of degenerate classical states as well as the semiclassical selection of states out of this manifold\cite{mulder10} are correctly captured within our approach.

We finally elaborate on intermediate spin magnitudes $S$. To the best of our knowledge there are only two previous systematic works to compare with, both investigating the $S=1$ case.\cite{gong15,li16} The most striking difference is that both studies find indications for a narrow non-magnetic phase around $g=0.3$ whereas our approach clearly detects magnetic order throughout the phase diagram at $S=1$. A possible reason for not finding this phase might be the neglect of three-particle vertices. However, such terms are subleading in $1/S$ and should quickly become irrelevant with increasing $S$. Conversely, if three-particle vertices were essential for the formation of a non-magnetic phase at $S=1$, their neglect would be even more severe for $S=1/2$. In the latter case, however, no systematic overestimation of long-range magnetic order is observed. Indeed, without our analysis of level repulsion terms (see Sec.~\ref{sc:local}) one would have rather guessed that magnetic order is underestimated within our spin-$S$ PFFRG method. This is because possible contributions from Hilbert-space sectors with smaller spin magnitudes could effectively increase quantum fluctuations. Our analysis in Sec.~\ref{sc:local}, however, points against such effects. For these reasons, we tend to believe in the accuracy of our results. 

Another difference is that for large enough $J_2$, Refs.~\onlinecite{gong15} and \onlinecite{li16} both find stripy order while we detect incommensurate spiral phases. The nature of the magnetic order in this regime has already been debated in the $S=1/2$ case where some methods identify spiral order\cite{oitmaa11,reuther11_2,zhang13} while others find a quantum locking of the magnetic wave vector at a high-symmetry point, yielding stripy order.\cite{albuquerque11,mezzacapo12,li12} Generally, with increasing $S$ the propensity for such type of quantum locking should become weaker, hence strengthening spiral order tendencies. We also note that the coupled cluster method applied in Ref.~\onlinecite{li16} did not probe the system with respect to spiral order. Furthermore, the DMRG studies in Ref.~\onlinecite{gong15} report conflicting spin patterns in this parameter regime when extrapolating the results to the thermodynamic limit. We therefore speculate that the restriction to small cylinder widths incompatible with incommensurate order might mask spiral order in DMRG. On the other hand, the PFFRG is not implemented on a finite cluster but only restricts the extent of the spin correlations. As a consequence, commensurate and incommensurate types of magnetism can both be described on equal footing within this approach. We should, however, also emphasize that we can generally not rule out the possibility that the neglected three-particle vertices realize such a quantum locking. We finally note that, since Refs.~\onlinecite{gong15} and \onlinecite{li16} might not have sufficiently taken into account the possibility of spiral spin configurations, this could also explain an erroneous detection of a magnetically disordered phase in a regime that is actually spiral ordered.

%====================================================================

%====================================================================
\section{Acknowledgements}
We thank C. Fr\"a\ss dorf, Y. Iqbal, R. Thomale, and M. Hering for stimulating discussions. J.R. is supported by the Freie Universit\"at Berlin within the Excellence Initiative of the German Research Foundation.
%====================================================================

\bibliography{ref}

\end{document}